\documentclass[journal]{IEEEtran}

\usepackage[utf8]{inputenc}
\usepackage[T1]{fontenc}
\usepackage{url}
\usepackage{ifthen}
\usepackage{cite}
\usepackage[cmex10]{amsmath}
\usepackage{stfloats}
\usepackage{graphicx}
\usepackage{epstopdf}
\usepackage{amsfonts}
\usepackage{enumitem}
\usepackage{amssymb}
\usepackage{mathrsfs}
\usepackage{bm}
\usepackage{mdwmath}
\usepackage{subfigure}
\usepackage{mdwtab}
\usepackage{dsfont}
\usepackage{color}
\usepackage{verbatim}
\usepackage{array}
\usepackage{epstopdf}
\usepackage{multirow}
\usepackage{booktabs}
\usepackage{lineno}
\usepackage{amsmath}
\usepackage{makecell}
\usepackage{algorithmic}
\usepackage{amsthm}

\usepackage{flushend}
\usepackage[ruled,vlined]{algorithm2e}

\DeclareMathOperator*{\argmin}{arg\,min}
\allowdisplaybreaks

\usepackage{titletoc}

\begin{document}
\title{Joint Optimization of Video-based AI Inference Tasks in MEC-assisted Augmented Reality Systems}

\author{Guangjin Pan, Heng Zhang, Shugong Xu,~\IEEEmembership{Fellow,~IEEE},\\  Shunqing Zhang,~\IEEEmembership{Senior Member,~IEEE}, and Xiaojing Chen
	\thanks{
		G. Pan, H. Zhang, S. Xu, S. Zhang and X. Chen are with Shanghai Institute for Advanced Communication and Data Science, Shanghai University, Shanghai 200444, China. Emails: \{guangjin\_pan, hengzhang, shugong, Shunqing, jodiechen\}@shu.edu.cn.

		Part of this work has been accepted by Globecom-2022. This work was supported in part by the National Natural Science Foundation of China (NSFC) under Grant 61871262, 62071284, and 61901251, the National Key R\&D Program of China grants 2017YFE0121400, 2019YFE0196600 and 2022YFB2902000, the Innovation Program of Shanghai Municipal Science and Technology Commission grants 20JC1416400 and 21ZR1422400, Pudong New Area Science \& Technology Development Fund, Key-Area Research and Development Program of Guangdong Province grant 2020B0101130012, Foshan Science and Technology Innovation Team Project grant FS0AA-KJ919-4402-0060, and research funds from Shanghai Institute for Advanced Communication and Data Science (SICS). The corresponding author is Shugong Xu.}
}
    \maketitle

\markboth{IEEE TRANSACTIONS ON COGNITIVE COMMUNICATIONS AND NETWORKING,~Vol.~XX, No.~XX, XXX~2022}%
{Shell \MakeLowercase{\textit{et al.}}: Bare Demo of IEEEtran.cls for IEEE Transactions on Magnetics Journals}
%



\IEEEtitleabstractindextext{%
\begin{abstract}
The high computational complexity and energy consumption of artificial intelligence (AI) algorithms hinder their application in augmented reality (AR) systems. However, mobile edge computing (MEC) makes it possible to solve this problem. This paper considers the scene of completing video-based AI inference tasks in the MEC system. We formulate a mixed-integer nonlinear programming problem (MINLP) to reduce inference delays, energy consumption and to improve recognition accuracy. We give a simplified expression of the inference complexity model and accuracy model through derivation and experimentation. The problem is then solved iteratively by using alternating optimization. Specifically, by assuming that the offloading decision is given, the problem is decoupled into two sub-problems, i.e., the resource allocation problem for the devices set that completes the inference tasks locally, and that for the devices set that offloads tasks. For the problem of offloading decision optimization, we propose a Channel-Aware heuristic algorithm. To further reduce the complexity, we propose an alternating direction method of multipliers (ADMM) based distributed algorithm. The ADMM-based algorithm has a low computational complexity that grows linearly with the number of devices. Numerical experiments show the effectiveness of proposed algorithms. The trade-off relationship between delay, energy consumption, and accuracy is also analyzed.
\end{abstract}

\begin{IEEEkeywords}
Mobile augmented reality, edge intelligence, mobile edge computing, resource allocation.
\end{IEEEkeywords}}

\maketitle

\IEEEdisplaynontitleabstractindextext

%
\IEEEpeerreviewmaketitle

\section{Introduction}
\IEEEPARstart{R}{ecently}, the development of networks, cloud computing, edge computing, artificial intelligence, and other technologies has triggered people's infinite imagination of the Metaverse \cite{Metaverse}. To enable users to interact between the real world and the virtual world, augmented reality (AR) technology plays a vital role. At the same time, artificial intelligence (AI), due to its learning and inference capabilities, has demonstrated a powerful ability in many fields such as automatic speech recognition (ASR) \cite{ASR}, natural language processing (NLP) \cite{NLP}, computer vision (CV) \cite{CV}, and so on. With the assistance of AI technology, AR can carry out deeper scene understanding and more immersive interactions.

However, the computational complexity of AI algorithms, especially deep neural networks (DNN), is usually very high. It is challenging to complete DNN inference timely and reliably on mobile devices with limited computation and energy capacity. In \cite{DeepSense}, experiments show that a typical single-frame image processing AI inference task takes about 600 ms even with speedup from the mobile GPU. In addition, continuously executing the above inference tasks can only last up to 2.5 hours on commodity devices. The above issues result in only a few AR applications currently using deep learning \cite{DeepDecision}. In order to reduce the inference time of DNNs, one way is to perform network pruning on the neural network \cite{pruning1,pruning3}. However, it could be destructive to the model if pruning too many channels, and it may not be possible to recover a satisfactory accuracy by fine-tuning \cite{pruning1}.

Edge AI \cite{EI1,EI2,EI3} is another approach to solving these problems. Integrating mobile edge computing (MEC) and AI technology has recently become a promising paradigm for supporting computationally intensive tasks. Edge AI transfers the inference and training process of AI models to the edge of the network close to the data source. Therefore, it can alleviate network traffic load, delay, and privacy problems. 

\subsection{Related Works}
Many existing studies use MEC's powerful computing capabilities to reduce delay \cite{Delay2}, energy consumption \cite{Energy2}, or both delay and energy consumption \cite{Delay-Energy3,Delay-Energy2,Delay-Energy4} through offloading. For example, \cite{Delay2} formulated an optimization problem aimed at minimizing the processing delay of eMBB and mMTC users by optimizing the users' transmit power in UAV-Assisted MEC systems. \cite{Energy2} develops a smart pricing mechanism to coordinate the computation offloading of multi-layer devices and reduces energy consumption. \cite{Delay-Energy3} uses the Stackelberg game method to optimize the task allocation coefficient, calculation resource allocation coefficient, and transmission power to minimize the energy consumption and delay of the NOMA-based MEC system.

For edge AI inference, existing research has made some progress. The authors in \cite{EI-energy1} propose a framework for jointly optimizing inference task selection and downlink coordinated beamforming to minimize communication power consumption in wireless networks. Similarly, \cite{EI-energy2} proposes an IRS-assisted edge inference system and designs a task selection strategy to minimize the energy consumption of uplink and downlink transmission and calculation. The work in \cite{EI-Delay-accuracy1} analyzes and models the transmission error probability, inference accuracy, and timeout probability of the AI-powered time-critical services. The work in \cite{EI-Delay1} uses a tandem queueing model to analyze queueing and processing delays of DL tasks in multiple DNN partitions. \cite{EI-Delay-energy1} joint optimizes the service placement, computational and radio resource allocation to minimize the users' total delay and energy consumption. \cite{pruning3} combines model pruning and DNN partitioning to achieve a 4.81x reduction on end-to-end delay. \cite{EI-accuracy2} designs the Edgent framework that can jointly optimize DNN partitioning and DNN right-sizing to maximize the inference accuracy while promising application delay requirements. These studies measure the inference time by experiments \cite{pruning3,EI-accuracy2} or assume that the inference task's computational complexity is proportional to the input data size but without derivation and proof \cite{EI-Delay1,EI-Delay-energy1}. However, these models of computational complexity are not rigorous enough or can not be generalized to different neural network models. 

As for the accuracy model, the authors in \cite{EI-Delay-accuracy2} designs an edge network orchestration algorithm named FACT, which boosts the performance of an edge-based AR system by optimizing the edge server assignment and video frame resolution selection for AR users. However, \cite{EI-Delay-accuracy2} builds an accuracy model by fitting an accuracy curve for specific tasks, which is not general. The work in \cite{EI-accuracy1} compresses image resolution locally and performs inference tasks on edge servers, aiming to maximize learning accuracy under constraints of delay and energy. \cite{EI-accuracy1} proposes using an abstract non-decreasing function to describe the relationship between accuracy and input image size, which cannot be used to analyze various AI inference tasks discriminately. Joint optimization is required when different tasks and models are jointly deployed. An insufficiently generalized accuracy model or an overly abstract model can adversely affect joint optimization. A general accuracy model is needed to measure various AI tasks.

\textcolor{black}{Among the above studies, most studies consider optimizing one or two performance metrics among the delay, energy consumption, and accuracy. The authors in \cite{EI-accuracy1} jointly considers delay, energy consumption and accuracy in image recognition scenarios. However, it aims at maximizing computational capacity under constraints of delay, energy consumption and accuracy, and the DNN model is only deployed in edge servers. In \cite{DeepDecision,EI-Delay-accuracy2,Trine}, video analytics scenarios are considered, but they do not jointly consider delay, energy and accuracy.}

\subsection{Contributions and Organizations}

\begin{figure}[tb]
\centering 
\includegraphics[height=2.4in,width=2.8in]{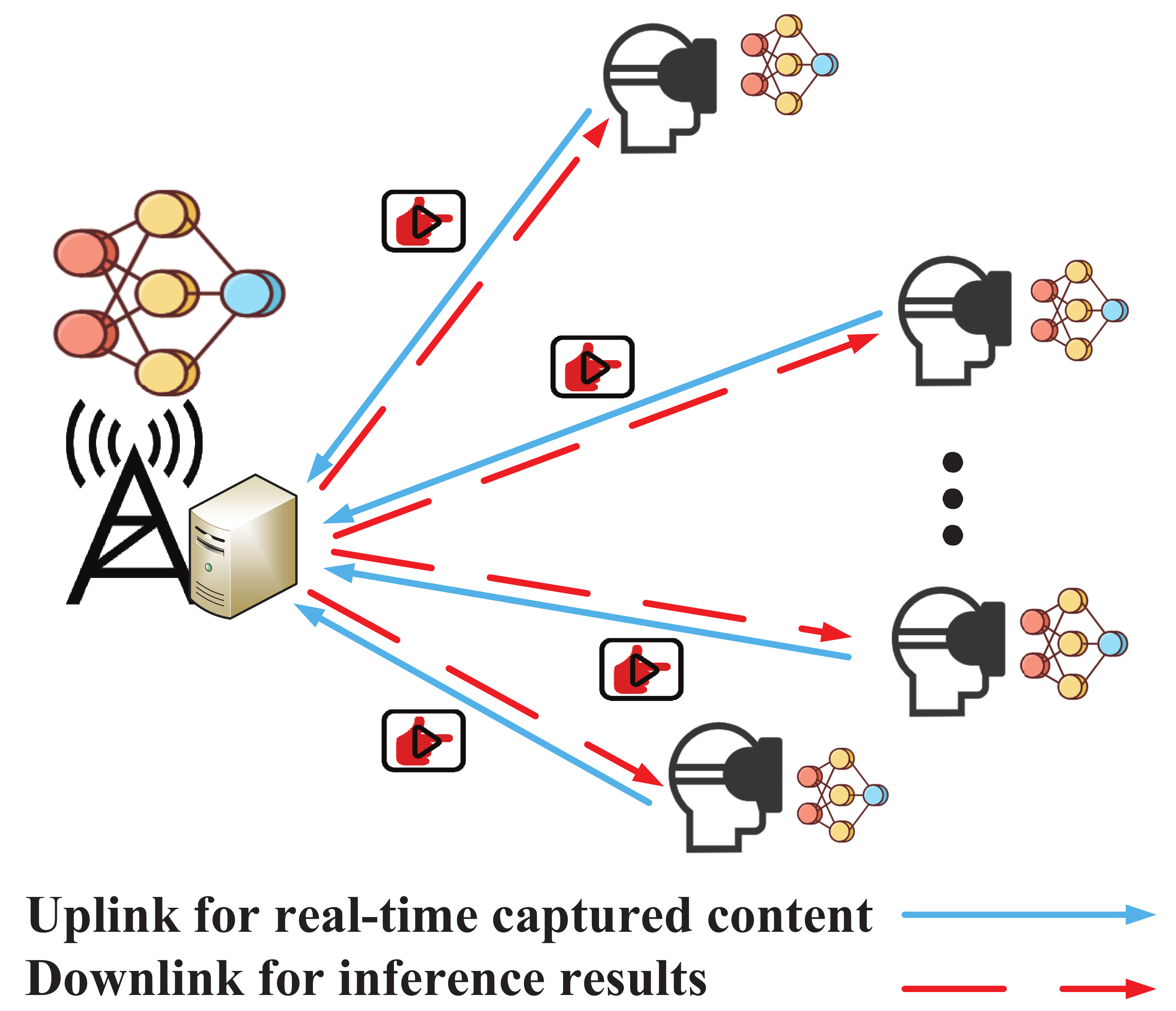} 
\caption{Multi-user MEC System model. The inference task can be executed on the local or the edge server. When the task is offloaded to the edge server, the uplink transmits the content captured in real-time, and the downlink transmits the inference result.}
\vspace{-4 mm}
\label{System} 
\end{figure}

In this paper, we consider a multi-user MEC system and assume that each device executes the video-based DNN inference task. Each device can be AR glasses, mobile robots, and so on. \textcolor{black}{In order to deepen AR's ability to understand the scene, we need to use time dimension information to improve perception. Therefore, we consider video-based application scenarios.for video-based AI inference tasks, there are two modes, e.g., frame-by-frame recognition mode (the input for each recognition is one frame) and multi-frame recognition mode (the input for each recognition is multiple frames). The frame-by-frame inference mode is used to deal with tasks with weak temporal correlation, such as face recognition and target tracking., and has been studied in \cite{DeepDecision,EI-Delay-accuracy2,Trine}. In this paper, we focus on multi-frame recognition tasks, such as gesture recognition and action recognition tasks. Since sampling in the spatial dimension brings extra computation \cite{EI-accuracy1}, we only sample in the temporal domain. At each inference, the device selects the most recent several frames from the history frames for transmission or inference.}

As shown in Fig. \ref{System}, mobile devices can transmit captured video to the edge server via wireless networks. The edge servers execute inference tasks and send results back to mobile devices. However, when communication and computing resources of the edge server are insufficient, devices can execute the inference task locally. We model the problem as a multi-objective optimization problem to optimize delay, energy consumption, and inference accuracy. The main contributions of this paper are summarized as follows,
\begin{itemize}
  \item{\textit{Multi-dimensional target optimization.}} High accuracy, low delay, and low energy consumption are indispensable for AR applications and must be optimized jointly. To explore the trade-off relationship between delay, energy, and accuracy, we formulate the video-based offloading problem as a mixed-integer nonlinear programming problem (MINLP), aiming to reduce service delays, energy consumption and improve recognition accuracy.
  \item{\textit{General computational complexity and accuracy models.}} To measure the computational complexity of neural network models with different architectures and different input sizes, we introduce the number of multiply-and-accumulate operations (MACs). We illustrate the main factors affecting DNN inference delay through experiments and show that MAC can be used as a good measure of the computational complexity of DNN inference tasks. We also propose a general model to represent the relationship between the inference accuracy and the number of input frames. This model is suitable for different video-based recognition tasks and different DNN architectures. We give simple expressions of the inference complexity and accuracy to simplify the optimization problem.
  \item{\textit{Channel-Aware scheduling scheme.}} To solve the optimization problem, we decompose the original problem. First, assuming that the offloading decision is given, we solve the resource allocation problems for the device set that completes the inference locally and the device set that offloads the tasks to the edge server, respectively. For edge DNN inference, we propose two algorithms based on search and geometric programming (GP) to solve the problem. Then, to obtain the optimal offloading policy, we propose a Channel-Aware heuristic algorithm. The original problem is solved iteratively through alternating optimization.
  \item{\textit{ADMM-based distributed resource allocation scheme.}} To avoid the high complexity of the heuristic algorithm, we propose an algorithm based on the Alternating direction method of multipliers (ADMM). The ADMM-based algorithm decomposes the original problem into parallel and tractable subproblems. Therefore, the total computational complexity of ADMM-based algorithms is more scalable than the heuristic algorithm, especially when the number of devices is large.
\end{itemize}

The rest of this paper is organized as follows. In Section \ref{Sec2}, we introduce system models, including delay, energy, and accuracy models. In Section \ref{Sec3}, we formulate the joint optimization problem and convert the original problem to a more tractable problem. Section \ref{Sec4} proposes a Channel-Aware heuristic algorithm to solve the proposed problem. In Section \ref{Sec5}, we propose another ADMM-based distributed resource allocation algorithm for the proposed problem, and analyze the computational complexity of the solution algorithm. Numerical results and analysis are presented in Section \ref{Sec6}. Finally, the paper is concluded in Section \ref{Sec7}.

\section{System Model} \label{Sec2}

In this section, we introduce a single-cell MEC system and establish delay, energy consumption, and accuracy models. As shown in Fig. \ref{System}, we consider a multi-user MEC system with one base station (BS) and $N$ mobile devices, denoted by the set $\mathcal{N}=\left\{1,2,\dots N\right\}$. Each device has a camera and needs to accomplish DNN inference tasks. Due to the limitation of device computational resources, DNN inference tasks can be placed on local or edge servers. The limited computational resource will lead to longer computing delay and greater power consumption when the inference task is executed locally. However, when the inference task is executed on the edge server, it will bring additional wireless transmission delay. In addition, accuracy is also a very important optimization target in DNN inference tasks.

\subsection{Offloading Framework}
\begin{figure}[tb]
\centering 
\includegraphics[height=2.5in,width=3in]{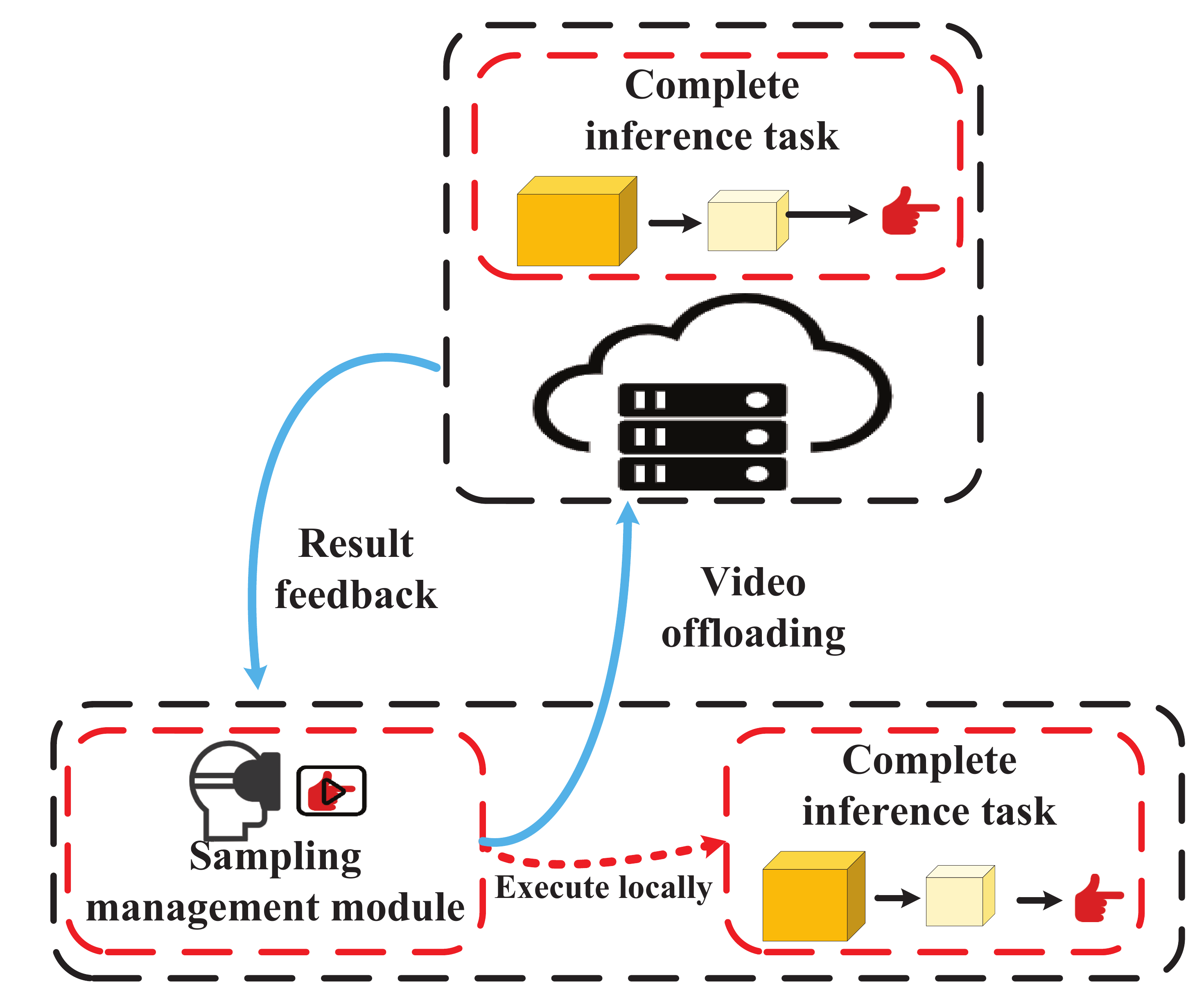} 
\caption{The overview of the video sampling and computing offloading system. The video sampling management module can control the sampling rate of the captured video and determine the number of video frames used for AI inference. Devices can transmit the video to the edge server or perform inference tasks locally based on the wireless channel information and computing capabilities.}
\label{overview} 
\vspace{-4 mm}
\end{figure}

In this paper, we only consider the binary offloading method. Binary offloading requires the DNN inference task to be fully executed either at the device or the MEC server. The overview of the DNN computing offloading system is depicted in Fig. \ref{overview}. First, devices sample the video captured in real-time in the temporal dimension to obtain a short video with a certain number of frames. Second, the DNN inference tasks are executed. These inference tasks can be executed locally on devices or the edge server. Therefore, each device's video sampling management module needs to select an appropriate video sampling rate (how many frames need to be input) and choose whether to offload the task to the MEC server. Denote $D_n$, $E_n$ and $\phi_n$ to be the total delay, energy consumption and recognition accuracy of the device $n$, respectively. The total delay and energy consumption of the device $n$ can be given by,
\begin{eqnarray} 
D_n  =  (1-x_n)D_n^{md}+x_n(D_n^t+D_n^{e}), \label{equ2-A-1}
\end{eqnarray}
\begin{eqnarray} 
E_n =  (1-x_n)E_n^{md}+x_nE_n^{t}, \label{equ2-A-2}
\end{eqnarray}
where $x_n$ indicates whether the inference task is executed on local or edge servers. $D_n^t$ is the transmission delay for uplink, $D_n^{md}$ is the local inference delay, and $D_n^{e}$ is the delay for completing inference at the edge server. $E_n^t$ and $E_n^{md}$ are the transmission and computational energy consumption, respectively. The delay and energy consumption for downloading computation results can be reasonably neglected because of the results' small data sizes.

\subsection{Delay and Energy Models for Inference}

The inference delay depends on the DNN model's architecture, the device's or server's computing power, and the input to the model. In this section, we first give a measure of the computational complexity of the DNN model and then give an expression for the inference delay and energy consumption.

Different AI recognition tasks may require different AI model architectures, including classic AI models such as Resnet-18, Resnet-34, Resnet-50, VGG-16, etc. \cite{Resnet18, vgg16}. In order to optimize AI inference tasks more reasonably, different AI models need a common method to evaluate computational complexity. In this paper, we use the number of MACs \cite{MAC_url} to measure the computational complexity of AI inference tasks. MACs calculation methods of layers (such as fully connected (FC) layers, convolutional layers and so on) can be obtained in \cite{MAC_url}. Taking 3D Convolutional Neural Network (3DCNN) as an example, the computational complexity (measured by MACs) of the $l^{th}$ layer of the $n^{th}$ device can be expressed as,
\begin{eqnarray}\label{equ2-B-1}
c_{n,l}= & o_{l} o_{l+1} \prod_{j=0}^2K_l^j, \prod_{j=0}^2M_{n,l+1}^j,
\end{eqnarray}
where $o_{l}$ is the number of input channels, $o_{l+1}$ is the number of output channels, $\prod_{j=0}^2K_l^j$ is the size of the convolution kernel, and $\prod_{j=0}^2M_{n,l+1}^j$ is the size of the output feature map. $j=0$ represents the temporal dimension (the number of frames), $j=1, 2$ represent spatial dimensions (pixels of one frame). Note that $o_l$, $o_{l+1}$, and $\prod_{j=0}^2K_l^j$ are all determined by the neural network architecture and $\prod_{j=0}^2M_{n,l+1}^j$ depends on the input size. The relation between the output feature size and the input size can be expressed as,
\begin{eqnarray}\label{equ2-B-2}
M_{n,l+1}^j=\frac{M_{n,l}^j-K_l^j+2d_l}{r_l}+1,
\end{eqnarray}
where ${r_l}$ is the stride and ${d_l}$ is the padding size. 

As mentioned above, the computational complexity of a DNN model is determined by the number of layers, the DNN model's architecture, and the input and output size. In this paper, we mainly focus on the impact of the number of input video frames $M_{n}$ on recognition accuracy and the allocation of communication and computing resources. The inference result will be more accurate with more frames $M_{n}$ input, but the communication and calculation overhead will be greater. The computational complexity of the $n^{th}$ device's task can be expressed as $C(M_{n})$. 

Then we give the expression for the inference delay and energy consumption. Denote $f^{max}$ and $f^{max}_n$ (in CPU cycle/s) to be the total computation resource of the edge server and mobile device $n$, respectively. Let $f^{e}_n$ and $f^{md}_n$ (in CPU cycle/s) denote the computation resource to device $n$ allocated by the edge server and the device, respectively. Therefore, the computing resources satisfy $\sum_{n \in \mathcal{N}} f_n^{e} \leq f^{max}$ and $ f_n^{md} \leq f^{max}_n$. The computation delay of the device $n$ and MEC can be respectively expressed as, \begin{eqnarray}
D_n^{md}=\frac{\rho C(M_{n})}{f_n^{md}}, \label{equ2-B-3}\\
D_n^{e}=\frac{\rho C(M_{n})}{f_n^{e}}, \label{equ2-B-4}
\end{eqnarray}
where $\rho$ (cycle/MAC) represents the number of CPU cycles required to complete a multiplication and addition, which depends on the CPU model.

As for energy consumption, denote $\kappa$ to be a coefficient determined by the corresponding device \cite{EI-accuracy1}, and the computational energy consumption of device $n$ can be expressed as,
\begin{eqnarray}
E_n^{md}=\kappa \rho C(M_{n}){f_n^{md}}^2. \label{equ2-B-5}
\end{eqnarray}

\subsection{Delay and Energy Models for Transmission}

We consider a time-division multiple access (TDMA) method for channel access. \textcolor{black}{ Specifically, each radio frame is divided into $N$ time slots for transmission, and each device can only transmit in its own time slot. We assume that the length of each radio frame is $\Delta T$, which is short enough (e.g., 10 ms in LTE or NR system \cite{EI-accuracy1}), and the length of a time slot is $\Delta Tt_n$.} 

Denote $h_n$ and $p_n$ to be the channel gain and transmission power of the device $n$, respectively. According to \cite{EI-Delay-energy1}, the achievable data rate of device $n$ can be expressed as, 
\begin{eqnarray}\label{equ2-C-1}
R_n=B_w log_2 \left( 1+\frac{p_nh_n}{B_w N_0} \right),
\end{eqnarray}
where $B_w$ and $N_0$ are the bandwidth and the variance of additive white Gaussian noise (AWGN), respectively. 

Let $d$ denote the data size of one video frame. Since we only want to analyze the impact of time dimension information (the number of input frames $M_n$) on recognition accuracy, $d$ is a constant value. \textcolor{black}{In each radio frame, the data size that can be transmitted is $\Delta TR_nt_n$. Therefore, for each transmission, $\lceil\frac{M_n d}{\Delta TR_nt_n}\rceil$ radio frames are required, where $\lceil \cdot \rceil $ means the ceil function. Considering that the length of the radio frame is much shorter than the transmission delay, the transmission delay for offloading to MEC can be written as,} 
\begin{eqnarray}\label{equ2-C-2}
D^t_n=\lceil\frac{M_n d}{\Delta TR_nt_n}\rceil \Delta T\approx\frac{M_n d}{R_nt_n},
\end{eqnarray}
where $t_n$ is the proportion of time that device n transmits. In addition, according to \cite{EI-accuracy1}, the energy consumption of each device to transmit its video can be expressed as,
\begin{eqnarray}\label{equ2-C-3}
E^t_n=\frac{M_n d}{R_n}p_n.
\end{eqnarray}

\subsection{Inference Tasks Accuracy Model}

As mentioned above, we mainly focus on the impact of the number of input video frames $M_{n}$ on recognition accuracy. We assume that the quality of the input video is the same for different devices. For a certain task and DNN model, the accuracy is only determined by the number of input frames. Therefore, the accuracy of device $n$ can be expressed as $\phi_n=\Phi({M_{n}})$. According to \cite{EI-energy-accuracy1}, more frames will lead to better inference accuracy, and as the input frames continue to increase, the performance gain will gradually decrease. Some prior studies also show that the relationship between frame rate and accuracy can be expressed as concave functions \cite{EI-Delay-accuracy2}. Therefore, we define $\Phi({M_{n}})$ as a monotone non-decreasing function to describe the relationship between the accuracy and the number of input frames.

\section{Problem Formulation}  \label{Sec3}
In this section, we formulate the optimization problem to reduce the system's delay and devices' energy consumption and improve accuracy. We analyze the difficulty of solving the problem. To simplify the problem, we make a reasonable conversion of the problem.

\subsection{Original Problem Formulation}
Based on the above analysis, combining \eqref{equ2-A-1}, \eqref{equ2-A-2}, \eqref{equ2-B-3}-\eqref{equ2-B-5}, \eqref{equ2-C-2}, and\eqref{equ2-C-3}, the $n^{th}$ device's delay and energy consumption can be expressed as,
\begin{eqnarray} 
D_n &= & (1-x_n) \frac{\rho C(M_{n})}{f_n^{md}}+x_n(\frac{\rho C(M_{n})}{f_n^{e}}+\frac{ M_{n}d}{R_nt_n}),\ \ \ \ \label{equ3-A-1}  \\
E_n  &= &(1-x_n) \kappa \rho C(M_{n}) {f_n^{md}}^2 +x_n(\frac{ M_{n}d}{R_n}p_n). \label{equ3-A-2}
\end{eqnarray}

Given the system model described previously, our goal is to reduce end-to-end delay and energy consumption and improve recognition accuracy. Each device follows the binary offloading policy. The mathematical optimization problem of the total cost (delay, energy consumption, and accuracy) can be expressed as,

Problem $\mathcal{P}1$ (\textit{Original Problem}):
\begin{align}
\mathop{\textrm{minimize}}_{\left\{ M_{n}, t_n, f_n^{md} , f_n^e, x_n\right\}}\  & \sum_{n \in \mathcal{N}} \bigg(\beta_1 D_n+\beta_2 E_n-  \beta_3\Phi(M_{n})\bigg),  \label{equ3-A-3} \\
\textrm{subject to} \ \ \ 
        & \Phi(M_{n}) \ge \alpha_n, \ \forall n \in \mathcal{N}, \tag{\ref{equ3-A-3}a}  \label{equ3-A-3a}\\
       & M_{n} \leq M^{max}_n, \ M_{n} \in \mathbb{Z}, \tag{\ref{equ3-A-3}b}  \label{equ3-A-3b}\\
       & \sum_{n\in \mathcal{N}} x_nt_n \le 1, \tag{\ref{equ3-A-3}c}  \label{equ3-A-3c}\\
       & \sum_{n\in \mathcal{N}}  x_nf_n^e \le f^{max} \tag{\ref{equ3-A-3}d},  \label{equ3-A-3d}\\
       & t_n, f_n^e \ge 0, \ \forall n \in \mathcal{N}, \tag{\ref{equ3-A-3}e}  \label{equ3-A-3e}\\
       & 0 \le f_n^{md} \le f_n^{max}, \forall n \in \mathcal{N}, \tag{\ref{equ3-A-3}f}  \label{equ3-A-3f}\\
       & x_n \in \left\{0,1\right\}, \forall n \in \mathcal{N},  \tag{\ref{equ3-A-3}g}  \label{equ3-A-3g} 
\end{align}
where $\alpha_n$ represents the recognition accuracy requirement, $\beta_1$, $\beta_2$, $\beta_3$ are the weight factors. \eqref{equ3-A-3a} represents the recognition accuracy requirement of each device. \eqref{equ3-A-3b} indicates the frame limit for the input video, $\mathbb{Z}$ is the set of integers, and $M^{max}_n$ is the maximum number of frames of the input video. \eqref{equ3-A-3c} and \eqref{equ3-A-3d} represent the communication and computation resource limitation, respectively. \eqref{equ3-A-3f} limits the computation resource of each device.

The optimization variables in original problem $\mathcal{P}1$ are the number of input video frames $M_n$, the proportion of transmission time $t_n$, the local computation resource $f_n^{md}$, the edge computation resource allocation $f_n^e$, and the offloading decision $x_n$. In addition, the first item in \eqref{equ3-A-3} is to reduce the total delay of computation and transmission, the second item is to reduce the device's energy consumption, and the last item is to improve the number of input video frames as well as the recognition accuracy because of the monotone non-decreasing function $\Phi(M_{n})$.

Problem $\mathcal{P}1$ is a non-convex MINLP problem and is difficult to be solved. First, the complexity function $C(M_n)$ is discrete and depends on the architecture of the DNN and the size of the input video. As the number of input frames $M_n$ increases, the computational complexity also increases. This kind of increase is irregular because it is affected by the structure of DNN layers, such as the stride and padding size of 3DCNN according to \eqref{equ2-B-2}. Therefore, $C(M_n)$ cannot be used for optimization directly. Second, as mentioned above, the accuracy function $\Phi(M_n)$ is non-decreasing. However, we cannot give a deterministic expression for $\Phi(M_n)$, so we can not optimize it. In addition, both $M_n$ and $x_n$ are integers, making the problem difficult to be solved.

\subsection{Problem Conversion}
To make the problem $\mathcal{P}1$ more tractable, we convert the problem. First, we give an approximate expression of the computational complexity function $C(M_n)$. According to \eqref{equ2-B-1} and \eqref{equ2-B-2}, the computational complexity of 3DCNN layers is proportional to the size of the input data. We can also obtain a similar conclusion in other types of layers, such as the FC layer \cite{MAC_url}. Based on the above conclusion and combined with the experiments in Sec. \ref{S5-1}, in order to simply express the computational complexity model, $C(M_{n})$ can be written as,
\begin{eqnarray}\label{equ3-B-1}
C(M_{n})=m_{c,0}M_{n}+m_{c,1} ,
\end{eqnarray}
where $m_{c,0} \ge 0$ and $m_{c,1}$ are constants and depend on the network model.

Second, we propose a general model to express the relationship between the accuracy and the number of input video frames. Considering that the function $\Phi(M_n)$ is monotonically non-decreasing and that as the number of input frames increases, the accuracy gain decreases, combining our experiments in Sec. \ref{S5-1}, we model function $\Phi({M_{n}})$ as,
\begin{eqnarray}\label{equ3-B-2}
\Phi(M_{n})=-\frac{m_{a,0}}{M_{n}+m_{a,1}}+m_{a,2},
\end{eqnarray}
where $m_{a,0} \ge 0$, $m_{a,2}  \ge 0$ and $m_{a,1}  > -1$ are constants and depend on the target of inference tasks and the architecture of DNN models.

Finally, we relax the range of the variable $M_{n}$. Considering that $\Phi(M_{n})$ is a monotone  non-decreasing function and depends on the recognition task and network architecture, in order not to lose generality, define $M_{n}^{min}=\argmin_{M_n}{\Phi(M_{n})}, \ \Phi(M_{n}) \ge \alpha_n, \ M_{n} \in \mathbb{Z}$. We can also relax $M_{n}$ into a closed connected subset of the real axis, and (\ref{equ3-A-3a}), (\ref{equ3-A-3b}) can be written as $M_{n} \in \left[ M^{min}_n,M^{max}_n \right]$. Then $[M_{n}]$ can be regarded as the number of input video frames, where $[ \cdot ]$ indicates rounding. We define two sets of devices, i.e. $\mathcal{N}_0=\{n\ | \ x_n=0, n \in \mathcal{N}\}$ and $\mathcal{N}_1=\{n \ | \ x_n=1, n \in \mathcal{N}\}$. $\mathcal{F}_{0,n}$ and $\mathcal{F}_{1,n}$ are the cost function of the device $n$ in sets $\mathcal{N}_0$ and $\mathcal{N}_1$, respectively. The problem $\mathcal{P}1$ can be rewritten as,

Problem $\mathcal{P}2$ (\textit{Converted Problem}):
\begin{align}
\mathop{\textrm{minimize}}_{\left\{ M_{n}, t_n, f_n^{md} ,f_n^e, x_n\right\}} \ \ &  \sum_{n \in \mathcal{N}_0}(1-x_n)\mathcal{F}_{0,n}({M_n},{f_n^{md}}) \nonumber \\ & +\sum_{n \in \mathcal{N}_1}x_n \mathcal{F}_{1,n}({M_n},{f_n^e},{t_n}),   \label{equ3-B-3} \\
\textrm{subject to} \  \  
       &M_{n} \in \left[ M^{min}_n,M^{max}_n \right], \tag{\ref{equ3-B-3}a}  \label{equ3-B-3a} \\
       &\eqref{equ3-A-3c}-\eqref{equ3-A-3g}, \nonumber
\end{align}
where 
\begin{align}
\mathcal{F}_{0,n}({M_n},{f_n^{md}}) = \ & \   \beta_1\frac{\rho C(M_{n})}{f_n^{md}}+\beta_2 \kappa\rho C(M_n)f_n^{md2} \nonumber\\
& -\beta_3\Phi(M_{n}),\label{equ3-B-4}\\
\mathcal{F}_{1,n}({M_n},{f_n^e},{t_n}) =\ & \  \beta_1 \frac{\rho C(M_{n})}{f_n^{e}}+ \beta_1\frac{M_nd}{R_nt_n} \nonumber\\
&+\beta_2 \frac{M_ndp_n}{R_n} -\beta_3\Phi(M_{n})   \label{equ3-B-5}.
\end{align}

\section{Optimization Problem Solving} \label{Sec4}

In this section, we decompose the problem $\mathcal{P}2$ and propose a Channel-Aware heuristic algorithm to solve it. First, supposing that the offloading decision (i.e., $\{x_n\}$) is given, we solve optimization problems for sets $\mathcal{N}_0$ and $\mathcal{N}_1$, respectively. Second, we propose a Channel-Aware heuristic algorithm to optimize the offloading decision $\{x_n\}$.

\subsection{Optimization Problem Solving for $\mathcal{N}_0$}

For set $\mathcal{N}_0$, i.e., when the device executes inference tasks locally, the optimization problem becomes,

Problem $\mathcal{P}_{\mathcal{N}_0}$ (\textit{Problem for $\mathcal{N}_0$}):
\begin{align}
 \mathop{\textrm{minimize}}_{\left\{ M_{n}, f_n^{md} \right\}} \ \  & \mathcal{F}_{\mathcal{P}_{\mathcal{N}_0}} \triangleq  \sum_{n \in \mathcal{N}_0} \mathcal{F}_{0,n}({M_n},{f_n^{md}}), \label{equ4-A-1} \\
\textrm{subject to} \ \ \  & \eqref{equ3-A-3f}, \  \eqref{equ3-B-3a}. \nonumber
\end{align}

The optimization variables in $\mathcal{P}_{\mathcal{N}_0}$ are the number of input video frames $M_n$ and the local computation resource $f_n^{md}$. Let $\{M_n^*, f_n^{md*}\}$ denote the optimal solution to $\mathcal{P}_{\mathcal{N}_0}$. We can derive the optimal solution to $\mathcal{P}_{\mathcal{N}_0}$ in a closed-form expression.

\textit{Theorem 1}: The optimal solution to $\mathcal{P}_{\mathcal{N}_0}$ is given by,
\begin{align}
f_n^{md*} = & \ \textrm{min}\{\sqrt[3]{(\frac{\beta_1}{2\beta_2\kappa})}, f_n^{max}\} \label{equ4-A-2}, \\
M_n^{*} = & \ \textrm{min} \{\textrm{max}\{ \sqrt{\frac{\beta_3 m_{a,0}}{\frac{ \beta_1 \rho m_{c,0}}{f_n^{md}}+\beta_2 \kappa\rho m_{c,0}f_n^{md2}}} \nonumber\\
& -m_{a,1}, M^{min}_n\},M^{max}_n\}. \label{equ4-A-3}
\end{align}

\textit{Proof:}  Please refer to Appendix A.

From \textit{Theorem 1}, we can see that the optimal
local CPU-cycle frequency $f_n^{md}$ is determined by the weight factors $\beta_1$, $\beta_2$, the coefficient of CPU energy consumption $\kappa$, and is limited by its corresponding upper bound $f_n^{max}$. More specifically, $f_n^{md}$ is proportional to $\beta_1^{\frac{1}{3}}$ and inversely proportional to $\beta_2^{\frac{1}{3}}$ and $\kappa^{\frac{1}{3}}$. As for the number of input video frames, when $\sqrt[3]{(\frac{\beta_1}{2\beta_2\kappa})} \le f_n^{max}$, combining \eqref{equ4-A-2} and \eqref{equ4-A-3}, we have,
\begin{align}
M_n^{*} = & \ \textrm{min} \{\textrm{max}\{ 3^{-\frac{1}{2}} 2^{\frac{1}{3}} \rho^{-\frac{1}{2}} \kappa^{-\frac{1}{6}} m_{c,0}^{-\frac{1}{2}} \beta_1^{-\frac{1}{3}} \beta_2^{-\frac{1}{6}} \beta_3^{\frac{1}{2}}m_{a,0}^{\frac{1}{2}}\nonumber\\
&  \ \ \ \  \ \ \ \  \ \ \ \ \ -m_{a,1}, M^{min}_n\},M^{max}_n\}. \label{equ4-A-4}
\end{align}
The optimization results corresponding to each device are only related to the parameters of the device itself and are not associated with the parameters of other devices.

\subsection{Optimization Problem Solving for $\mathcal{N}_1$}
Then we solve the optimization problem of $\mathcal{N}_1$. The problem $\mathcal{P}2$ can be written as,

Problem $\mathcal{P}_{\mathcal{N}_1}$ (\textit{Problem for $\mathcal{N}_1$}):
\begin{align}
 \mathop{\textrm{minimize}}_{\left\{{M_n},{f_n^e},{t_n} \right\}} \ \ \ & \sum_{n \in \mathcal{N}_1} \mathcal{F}_{1,n}({M_n},{f_n^e},{t_n}) ,   \label{equ4-B-1} \\
\textrm{subject to} \ \ \ \ & \eqref{equ3-A-3c}, \ \eqref{equ3-A-3d}, \ \eqref{equ3-A-3e}, \  \eqref{equ3-B-3a}. \nonumber
\end{align}

The optimization variables in the the problem $\mathcal{P}_{\mathcal{N}_1}$ are the number of input video frames $M_n$, the edge computation resource $f_n^e$, and the proportion of transmission time $t_n$. Let $\{M_n^*, f_n^{e*}, t_n^{*}\}$ denote the optimal solution to $\mathcal{P}_{\mathcal{N}_1}$. We can obtain the optimal solution to $\mathcal{P}_{\mathcal{N}_1}$ using the method of Lagrange multiplier. The partial Lagrangian function can be written as,
\begin{align}
\! \mathcal{L}_{\mathcal{P}_{\mathcal{N}_1}} \! \! =  \! & \! \sum_{n \in \mathcal{N}_1} \! \! \left( \frac{\beta_1\rho C(M_{n})}{f_n^{e}}+ \frac{\beta_1M_nd}{R_nt_n}
+\frac{\beta_2 M_ndp_n}{R_n} - \beta_3\Phi(M_{n})  \right) \nonumber\\ & +\mu_0 (\sum_{n\in \mathcal{N}_1} t_n - 1)  + \mu_1 (\sum_{n\in \mathcal{N}_1} f_n^e - f^{max}) , \label{equ4-B-2} 
\end{align}

First of all, according to \eqref{equ4-B-2}, supposing that $M_n^{*}$ is given, we can solve the problem $\mathcal{P}_{\mathcal{N}_1}$ based on the Karush-Kuhn-Tucker (KKT) condition. We can obtain the function expressions of $f_n^{e*}$ and $t_n^{*}$ relative to $M_n$, as shown in the following theorem.

\textit{Theorem 2}: The function expressions of $f_n^{e*}$ and $t_n^{*}$ relative to $M_n^{*}$ are given by,
\begin{align}
f_n^{e*} = & \ \frac{f^{max} \sqrt{C(M_n^{*})}}{\sum\limits_{i\in \mathcal{N}_1} \sqrt{C(M_i^{*})}}, \label{equ4-B-3}   \\
t_n^{*} = & \ \frac{\sqrt{\frac{M_n^{*}}{R_n}}}{\sum\limits_{i\in \mathcal{N}_1} \sqrt{\frac{M_i^{*}}{R_i}}}. \label{equ4-B-4} 
\end{align}

\textit{Proof:}  Please refer to Appendix B.

Combining \eqref{equ4-B-1}, \eqref{equ4-B-3} and \eqref{equ4-B-4}, the problem $\mathcal{P}_{\mathcal{N}_1}$ can be written as an optimized function containing only the variable $M_n$ as follows,

\begin{algorithm}[t]
	\caption{Algorithm 1: Search-Based Algorithm for solving $\mathcal{P}_{\mathcal{N}_1}$}
	\label{algorithm1}
	\KwIn{ The offloading policy $\mathcal{N}_1$, the channel gain $\{h_n\}$, and other system parameters.}
	\KwOut{ $\{M_n^\star,f_n^{e \star},t_n^{\star}\}$}
	\BlankLine
    Initialize the result of cost function $\mathcal{F}_{{\widetilde{\mathcal{P}_{\mathcal{N}_1}}}}^{\star}$ to a sufficiently large value;
    
    Calculate the achievable data rate $\{R_n\}$ using \eqref{equ2-C-1};
    
    \ForEach{$\{ M_n \}  \in \mathcal{M} $}
    { 
     Compute $\mathcal{F}_{{\widetilde{\mathcal{P}_{\mathcal{N}_1}}}}$ using \eqref{equ4-B-5};\\
     \If{
        $\mathcal{F}_{{\widetilde{\mathcal{P}_{\mathcal{N}_1}}}}<\mathcal{F}_{{\widetilde{\mathcal{P}_{\mathcal{N}_1}}}}^{\star}$
     }{
        $\{ M_n^\star \}=\{ M_n\} $; \
        $\mathcal{F}_{{\widetilde{\mathcal{P}_{\mathcal{N}_1}}}}^{\star}=\mathcal{F}_{{\widetilde{\mathcal{P}_{\mathcal{N}_1}}}}$;
    }}
    
    Calculate $\{f_n^{e \star}\}$ and $\{t_n^{\star}\}$ using \eqref{equ4-B-3} and \eqref{equ4-B-4};
    \BlankLine
    \Return $\{M_n^\star\}$, $\{f_n^{e \star}\}$, and $\{t_n^{\star}\}$.
    
\end{algorithm}

Problem ${\widetilde{\mathcal{P}_{\mathcal{N}_1}}} $ (\textit{${M_n}$ Optimization Problem for $\mathcal{N}_1$ }):
\begin{align}
\mathop{\textrm{minimize}}_{\left\{{M_n} \right\}} \ \   \mathcal{F}_{{\widetilde{\mathcal{P}_{\mathcal{N}_1}}}} & \triangleq  \frac{\beta_1\rho}{f^{max}}{ (\sum\limits_{n\in \mathcal{N}_1} \sqrt{C(M_n)})^2} \nonumber \\ 
& +{\beta_1 d (\sum\limits_{n\in \mathcal{N}_1} \sqrt{\frac{M_n}{R_n}})^2}  + \beta_2 d p_n (\sum\limits_{n\in \mathcal{N}_1} \frac{M_n}{R_n}) 
\nonumber \\ & - \sum\limits_{n\in \mathcal{N}_1} \beta_3\Phi(M_{n}), \label{equ4-B-5} \\
\textrm{subject to} \ \ \ \ & \eqref{equ3-B-3a}.  \nonumber
\end{align}

Denote $\mathcal{M}_n^{opt}=\{M_n \ |\ M_n^{min} \le M_n \le M_n^{max}, M_n \in  \mathbb{Z} \}$ to be the optional video frame number of device $n$. The optimal solution can be obtained by searching for $\{M_n\} \in \mathcal{M} $, where $\mathcal{M} = \{\{M_i\}\ |\ M_i \in \mathcal{M}_i^{opt}, i \in \mathcal{N}_1\}$. The detail of the search based algorithm is shown in Algorithm \ref{algorithm1}.

\begin{algorithm}[t]
	\caption{Algorithm 2: GP-Based Algorithm for solving $\mathcal{P}_{\mathcal{N}_1}$}
	\label{algorithm2}
	\KwIn{ The offloading policy $\mathcal{N}_1$, the channel gain $\{h_n\}$, and other system parameters.}
	\KwOut{ $\{M_n^\star,f_n^{e \star},t_n^{\star}\}$}
	\BlankLine

    Calculate the achievable data rate $\{R_n\}$ using \eqref{equ2-C-1};
    
    Use the CVX tool to solve \eqref{equ4-B-7} and get $\{\hat{M_n^\star\}}$;
    
    $ \{ M_n^\star\}= \{ [e^{\hat{M_n^\star}}] \}$;
    
    Calculate $\{f_n^{e \star}\}$ and $\{t_n^{\star}\}$ using \eqref{equ4-B-3} and \eqref{equ4-B-4};
    \BlankLine
    \Return $\{M_n^\star\}$, $\{f_n^{e \star}\}$, and $\{t_n^{\star}\}$.
    
\end{algorithm}

\textcolor{black}{Considering that the problem $\mathcal{P}_{\mathcal{N}_1}$ is convex when $M_n$ is given, Algorithm \ref{algorithm1} is global optimal. However, When the number of devices grows large, the computational complexity of the Search-based algorithm will become very high or even unacceptable. In this paper, we also propose a GP-based sub-optimal algorithm to solve the problem $\mathcal{P}_{\mathcal{N}_1}$.} First, we relax the objective function of the problem $\mathcal{P}_{\mathcal{N}1}$. We introduce the function,
$\widehat{\Phi}(M_{n})=-\frac{m_{a,0}}{M_{n}}+m_{a,2}$, 
and $\mathcal{P}_{\mathcal{N}_1}$ can be rewritten as,

Problem ${{\mathcal{P}_{GP_{\mathcal{N}_1}}}} $ (\textit{GP-based Problem for $\mathcal{N}_1$}):
\begin{align}  
\mathop{\textrm{minimize}}_{\left\{{M_n},{f_n^e},{t_n} \right\}} \ \ \ & \sum_{n \in \mathcal{N}_1} \bigg( \beta_1 \frac{\rho C(M_{n})}{f_n^{e}}+ \beta_1\frac{M_nd}{R_nt_n}
\nonumber \\
& +\beta_2 \frac{M_ndp_n}{R_n} -  \beta_3\widehat{\Phi}(M_{n})  \bigg) ,   \label{equ4-B-6} \\
\textrm{subject to} \ \ \ \ & \eqref{equ3-A-3c}, \ \eqref{equ3-A-3d}, \ \eqref{equ3-A-3e}, \  \eqref{equ3-B-3a}. \nonumber
\end{align}
It is a non-convex GP problem. Inspired by \cite{convex}, the GP problem can be transformed into a convex problem by changing variables and transforming the objective and constraints. Therefore, introducing variables, $\hat{M_{n}}=\ln{{M_{n}}}, \hat{f_n^{e}}=\ln{f_n^{e}},\hat{t_n}=\ln{t_n} $, and the problem can be written as,

Problem ${\widetilde{\mathcal{P}_{GP_{\mathcal{N}_1}}}} $ (\textit{Converted GP-based Problem for $\mathcal{N}_1$}):
\begin{align}
\! \! \! \! \! \! \mathop{\textrm{minimize}}_{\left\{ \hat{M_{n}}, \hat{t_n} ,\hat{f_n^e}\right\}}   & \!  \sum_{n \in \mathcal{N}_1} \bigg( \beta_1 \rho m_{c,0} e^{\hat{M_n}-\hat{f_n^e}}  + \beta_1\rho m_{c,1} e^{-\hat{f_n^e}}   \nonumber \\ 
\! & \! \! + \! \frac{\beta_1 d e^{\hat{M_n}-\hat{t_n}}}{R_n} \! + \! \frac{\beta_2 d p_n e^{ \hat{M_n}}}{R_n} \! +  \! \beta_3 m_{a,0} e^{-\hat{M_n}} \! \bigg),    \label{equ4-B-7} \\
\textrm{subject to}   
        & \ \ \hat{M_{n}} \in \left[ \ln M_n^{min},\ln M_n^{max} \right], \forall n \in \mathcal{N}_1, \tag{\ref{equ4-B-7}a}  \label{equ4-B-7a}\\
       & \sum_{n\in \mathcal{N}_1} x_ne^{\hat{t_n}} \le 1, \tag{\ref{equ4-B-7}b}  \label{equ4-B-7b}\\
       & \sum_{n\in \mathcal{N}_1}  x_ne^{\hat{f_n^e}} \le f^{max}, \tag{\ref{equ4-B-7}c}  \label{equ4-B-7c}
\end{align}
which is strictly convex problem that can be solved using the CVX tool \cite{cvx}. Considering that $M_n$ is an integer, the result of CVX optimization needs to be post-processed. Details of the GP-based algorithm are shown in Algorithm \ref{algorithm2}.

\subsection{Optimization of Offloading Policy $\{x_n\}$}

\begin{algorithm}[t]
	\caption{Algorithm 3: Channel-Aware heuristic algorithm for Optimizing Offloading Policy $\{x_n\}$}
	\label{algorithm3}
	\KwIn{Parameters corresponding to the problem $\mathcal{P}1$.}
	\KwOut{Offloading policy $\mathcal{N}_0$ and $\mathcal{N}_1$.}
	\BlankLine

    Calculate the cost function $\{\mathcal{F}_{0,n}\}$ for the set $\mathcal{N}$ using \eqref{equ4-A-2} and \eqref{equ4-A-3} ;
    
    Set $\mathcal{N}_0=\emptyset$, $\mathcal{N}_1=\mathcal{N}$;
    
    Calculate the cost function $\{\mathcal{F}_{1,n}\}$ corresponding to the set $\mathcal{N}_1$ using Algorithm \ref{algorithm1} or Algorithm \ref{algorithm2};
    
    Set $Flag=1$;
    
    \While{$Flag==1$}{

    $k=\textrm{argmin}_n h_n, n \in \mathcal{N}_1$;
    
    $\mathcal{N}_0^*=\mathcal{N}_0 \cup \{k\}$, $\mathcal{N}_1^*=\mathcal{N}_1-\{k\}$;
    
    Calculate the cost function $\{\mathcal{F}_{1,n}^*\}$ corresponding to the set $\mathcal{N}_1^*$ using Algorithm \ref{algorithm1} or Algorithm \ref{algorithm2};\\
    \If{$\sum_{n \in \mathcal{N}_0} \mathcal{F}_{0,n} + \sum_{n \in \mathcal{N}_1} \mathcal{F}_{1,n} > \sum_{n \in \mathcal{N}_0^*} \mathcal{F}_{0,n}+\sum_{n \in \mathcal{N}_1^*} \mathcal{F}_{1,n}^* $}
    {
    $\mathcal{F}_{1,n}=\mathcal{F}_{1,n}^*, \forall n\in \mathcal{N}_1^*$;\\
    $\mathcal{N}_0=\mathcal{N}_0^*$;
    $\mathcal{N}_1=\mathcal{N}_1^*$;\\
    }\Else{
    $Flag=0$;
    }
    }
    \Return $\mathcal{N}_0$ and $\mathcal{N}_1$.
    
\end{algorithm}

Considering the complexity of Search-based offloading policy algorithm becomes high when the number of devices $N$ grows large. In this section, we propose a Channel-Aware heuristic algorithm to optimize the offloading decision $\{x_n\}$. Inspired by the \textit{Theorem 1} and \textit{Theorem 2}, when executing inference locally, the cost function $\mathcal{F}_{0,n}$ and optimization variables $f_n^{md}$, $M_n$ only depend on the device's own parameters. However, for edge set $\mathcal{N}_1$, the cost function is related to the number and parameters of devices in the set $\mathcal{N}_1$. The Channel-Aware heuristic algorithm is shown in Algorithm \ref{algorithm3}. First, calculate the cost function $\{\mathcal{F}_{0,n}\}$ of set $\mathcal{N}_0$ when each device's task is executed locally. Second, assuming that all devices are offloaded to the edge server for inference and $|\mathcal{N}_1|=N$. In each iteration, the cost function $\{\mathcal{F}_{1,n}\}$ corresponding to each device of $\mathcal{N}_1$ is obtained. We select the device $k$ with smallest channel gain in set $\mathcal{N}_1$. Try to put the device $k$ from the set $\mathcal{N}_1$ into the set $\mathcal{N}_0$ and compute the cost of new sets. If the total cost of new sets is reduced, continue the next iteration. Otherwise, put the device $k$ back to the set $\mathcal{N}_1$.

\section{JOINT OPTIMIZATION USING ADMM-BASED Method} \label{Sec5}

The complexity of the Channel-Aware heuristic algorithm becomes high when the number of UE grows. In this section, We propose an ADMM-based algorithm. The ADMM-based algorithm can decompose $\mathcal{P}2$ into $N$ parallel sub-problems. Each user only needs to solve one sub-problem, and the average complexity of each device will be reduced.

\subsection{ADMM-based Problem Conversion}
To make the original problem tractable, we jointly consider the problem ${{\mathcal{P}_{2}}}$ and problem ${\widetilde{\mathcal{P}_{GP_{\mathcal{N}_1}}}}$, and we converted the problem into a GP-based problem,

Problem $\mathcal{P}_3$ (\textit{Converted GP-based Problem}):
\begin{align}
\mathop{\textrm{minimize}}_{\left\{ \hat{M_{n}}, \hat{t_n}, \hat{f_n^{md}} ,\hat{f_n^e}, x_n\right\}} \ \ &  \sum_{n \in \mathcal{N}}\bigg[(1-x_n)\hat{\mathcal{F}_{0,n}}(\hat{M_n},\hat{f_n^{md}}) \nonumber \\ & +x_n \hat{\mathcal{F}_{1,n}}(\hat{M_n},\hat{f_n^e},\hat{t_n})\bigg],   \label{equ5-A-1} \\
\textrm{subject to} \  \ 
        & \hat{f_n^{md}} \le \ln f_n^{max}, \forall n \in \mathcal{N}, \tag{\ref{equ5-A-1}a}  \label{equ5-A-1a} \\
        &\eqref{equ3-A-3g}, \eqref{equ4-B-7a}-\eqref{equ4-B-7c}, \nonumber 
\end{align}
where $\hat{M_{n}}=\ln{{M_{n}}}$, $\hat{f_n^{md}}=\ln{f_n^{md}}$, $ \hat{f_n^{e}}=\ln{f_n^{e}}$, and $\hat{t_n}=\ln{t_n} $. $\hat{\mathcal{F}_{0,n}}(\hat{M_n},\hat{f_n^{md}})$ and $\hat{\mathcal{F}_{1,n}}(\hat{M_n},\hat{f_n^e},\hat{t_n})$ are given by,
\begin{align}
\hat{\mathcal{F}_{0,n}}(\hat{M_n},\hat{f_n^{md}}) = \ & \  \beta_1 \rho m_{c,0} e^{\hat{M_n}-\hat{f_n^{md}}}  + \beta_1\rho m_{c,1} e^{-\hat{f_n^{md}}} 
\nonumber \\
&+ \beta_2\kappa  m_{c,0} e^{\hat{M_n}+2\hat{f_n^{md}}} \nonumber \\
&+ \beta_2\kappa  m_{c,1} e^{2\hat{f_n^{md}}} +\beta_3 m_{a,0} e^{-\hat{M_n}}, \label{equ5-A-2}\\
\hat{\mathcal{F}_{1,n}}(\hat{M_n},\hat{f_n^e},\hat{t_n}) =\ & \  \beta_1 \rho m_{c,0} e^{\hat{M_n}-\hat{f_n^e}}  + \beta_1\rho m_{c,1} e^{-\hat{f_n^e}} 
\nonumber \\
&+ \frac{\beta_1 d e^{\hat{M_n}-\hat{t_n}}}{R_n} + \frac{\beta_2 d p_n e^{ \hat{M_n}}}{R_n} \nonumber \\
& +\beta_3 m_{a,0} e^{-\hat{M_n}}, \label{equ5-A-3} 
\end{align}

The optimization variables $\{\hat{t_n},\hat{f_n^e}\}$ are coupled among the devices in the constraints \eqref{equ4-B-7b} and \eqref{equ4-B-7c}. To decompose the problem $\mathcal{P}_3$, we introduce local variables $\{y_n\}$ and $\{z_n\}$. Then, the ADMM-based problem can be written as,

Problem $\mathcal{P}_{4}$ (\textit{ADMM-based Problem}):
\begin{align}
\mathop{\textrm{minimize}}_{\left\{ \hat{M_{n}}, \hat{t_n}, \hat{f_n^{md}} ,\hat{f_n^e}, x_n, y_n, z_n\right\}} \ \ &    \sum_{n \in \mathcal{N}}  \hat{\mathcal{F}_{n}}(x_n,\hat{M_n},\hat{f_n^{md}},y_n,z_n)  \nonumber \\ 
& +  {g}(\hat{f_n^e},\hat{t_n}),\label{equ5-A-4} \\
\textrm{subject to} \  \  
    & y_n=\hat{f_n^{e}}, z_n=\hat{t_n}, \tag{\ref{equ5-A-4}a}  \label{equ5-A-4a} \\ 
    &\eqref{equ3-A-3g},  \eqref{equ4-B-7a}, \eqref{equ5-A-1a}, \nonumber
\end{align}
where,
\begin{align}
&\hat{\mathcal{F}_{n}}(x_n,\hat{M_n},\hat{f_n^{md}},y_n,z_n)= (1-x_n)\hat{\mathcal{F}_{0,n}}(\hat{M_n},\hat{f_n^{md}}) \nonumber \\
& \qquad \qquad \qquad \qquad \qquad \quad \ +x_n \hat{\mathcal{F}_{1,n}}(\hat{M_n},x_n,y_n), \label{equ5-A-5}\\
&{g}(\hat{f_n^e}, \hat{t_n}) =  \left\{
{\begin{aligned}
0, \qquad     \mathop{\textrm{if}} (\hat{f_n^e},\ \hat{t_n}) \in \mathcal{G},\\
+\infty \qquad \quad, \mathop{\textrm{otherwise}}  ,
\end{aligned}} \right. \label{equ5-A-6}
\end{align}
and,
\begin{align}
\! \! \! \! \mathcal{G}=\bigg\{(\hat{f_n^e},\ \hat{t_n})|\sum_{n\in \mathcal{N}_1} x_n e^{\hat{t_n}} \le 1,\sum_{n\in \mathcal{N}_1}  x_n e^{\hat{f_n^e}} \le f^{max}\bigg\}. \label{equ5-A-7}
\end{align}

\subsection{ADMM-based Problem Solving}

The problem $\mathcal{P}_{4}$ can be effectively solved using the ADMM algorithm. We can write a partial augmented Lagrangian of the problem $\mathcal{P}_{4}$ as,
\begin{align}
\mathcal{L}_4(\bm{u},\bm{v},\bm{\theta})&=\sum_{n \in \mathcal{N}}  \hat{\mathcal{F}_{n}}(x_n,\hat{M_n},\hat{f_n^{md}},y_n,z_n) +  {g}(\hat{f_n^e},\hat{t_n}) \nonumber \\
&+ \sum_{n \in \mathcal{N}}\theta^f_n(y_n-\hat{f_n^e})+ \sum_{n \in \mathcal{N}}\theta^t_n(z_n-\hat{t_n}) \nonumber \\
& +\sum_{n \in \mathcal{N}}\frac{s}{2}(y_n-\hat{f_n^e})^2 +\sum_{n \in \mathcal{N}}\frac{s}{2}(z_n-\hat{t_n})^2, \label{equ5-B-1}
\end{align}
where $\bm{u}=\{x_n,\hat{M_n},\hat{f_n^{md}},y_n,z_n\}$, $\bm{v}=\{\hat{f_n^e},\hat{t_n}\}$, $\bm{\theta}=\{\theta^f_n,\theta^t_n\}$, and $s$ is a fixed step size. Therefore, the dual function is,
\begin{align}
p(\bm{\theta}) =& \mathop{\textrm{minimize}}_{\bm{u},\bm{v}} \mathcal{L}_4(\bm{u},\bm{v},\bm{\theta}) \label{equ5-B-2} \\
\textrm{subject to} \ &  \eqref{equ3-A-3g}, \eqref{equ4-B-7a} , \eqref{equ5-A-1a}, \nonumber\  \ 
\end{align}
and the dual problem can be given by,
\begin{align}
\mathop{\textrm{maximize}}_{\bm{\theta}} p(\bm{\theta}),  \label{equ5-B-3}\ 
\end{align}

The problem \eqref{equ5-B-2} can be solved by iteratively updating $\bm{u}$, $\bm{v}$, and $\bm{\theta}$ \cite{ADMM}. Let $\{\bm{u}^i,\bm{v}^i,\bm{\theta}^i\}$ denote the values in the $i^{th}$ iteration. In the $i^{th}$ iteration, the update strategies of the variables are as follows,

\subsubsection{Step 1}Local variables update. In this step, we first update the local variables $\bm{u}$. Given variable $\bm{v}^i$ and $\bm{\theta}^i$, we minimize  $\mathcal{L}_4(\bm{u},\bm{v},\bm{\theta})$ by,
\begin{align}
\bm{u}^{i+1} =& \mathop{\textrm{argminimize}}_{\bm{u}} \mathcal{L}_4(\bm{u},\bm{v}^i,\bm{\theta}^i). \label{equ5-B-4}
\end{align}
The problem \eqref{equ5-B-3} can be decomposed into $N$ parallel subproblems. For each subproblem, we consider two cases where $x_n=0$ and $x_n=1$, and express the problem as, 

\begin{align}
\! \! \! \! \! \left\{
{\begin{aligned}
\! \! \mathop{\textrm{minimize}}_{\{\hat{M_n},\hat{f_n^{md}},y_n,z_n\}} \! \! \hat{\mathcal{F}_{0,n}}(\hat{M_n},\hat{f_n^{md}})  = \! \theta^f_n y_n+ \sum_{n \in \mathcal{N}}\frac{s}{2}(y_n-\hat{f_n^e})^2 \\  + \theta^t_n z_n +\sum_{n \in \mathcal{N}}\frac{s}{2}(z_n-\hat{t_n})^2,   \qquad \quad    \mathop{\textrm{if}} x_n=0,  \\
\! \! \mathop{\textrm{minimize}}_{\{\hat{M_n},y_n,z_n\}}  \hat{\mathcal{F}_{1,n}}(\hat{M_n},y_n,z_n) \! = \! \theta^f_n y_n + \sum_{n \in \mathcal{N}}\frac{s}{2}(y_n-\hat{f_n^e})^2 \\+ \theta^t_n z_n +\sum_{n \in \mathcal{N}}\frac{s}{2}(z_n-\hat{t_n})^2,  \qquad  \quad \mathop{\textrm{if}} x_n=1. 
\end{aligned}} \right. \label{equ5-B-5}
\end{align}
These problems are both strictly convex problems that can be solved using the CVX tool \cite{cvx}. Therefore, we can calculate the objective value for $x_n=0$ and $x_n=1$ and choose the smaller one as the final result. After solving $N$ parallel subproblems, the optimal solution to \eqref{equ5-B-4} is given by $\bm{u}^{i+1}=\{(x_n)^{i+1},(\hat{M_n})^{i+1},(\hat{f_n^{md}})^{i+1},(y_n)^{i+1},(z_n)^{i+1}\}$.


\subsubsection{Step 2}Global variables update. In the second step, we update the global variables $\bm{v}$. By the definition of $g(\bm{v})$ in \eqref{equ5-A-6}, $ \bm{v}^{i+1} \in \mathcal{G}$ must hold at the optimum. Therefore, the subproblem can be equivalently written as,
\begin{align}
\bm{v}^{i+1} =& \mathop{\textrm{argminimize}}_{\{\hat{f_n^e},\hat{t_n}\}} \sum_{n \in \mathcal{N}}(\theta^f_n)^i(-\hat{f_n^e})+ \sum_{n \in \mathcal{N}}(\theta^t_n)^i(-\hat{t_n}) \nonumber \\
& +\sum_{n \in \mathcal{N}}\frac{s}{2}(y_n^{i+1}-\hat{f_n^e})^2 +\sum_{n \in \mathcal{N}}\frac{s}{2}(z_n^{i+1}-\hat{t_n})^2, \label{equ5-B-6}\\
 & \qquad \textrm{subject to}, \quad   \eqref{equ4-B-7b}, \eqref{equ4-B-7c}. \nonumber\  \ 
\end{align}

The problem can also be solved by the CVX tool \cite{cvx}. We propose a low-complexity scheme to solve this subproblem. Considering the constraints \eqref{equ4-B-7b} and \eqref{equ4-B-7c}, let $\mu_f$ and $\mu_t$ denote the Lagrangian multipliers. The closed-form optimal solution of this subproblem can be expressed as,
\begin{align}
(\hat{f_n^e})^{i+1} =& {y_n}^{i+1}+\frac{(\theta^f_n)^i-\mu_f}{s}, \label{equ5-B-7}\\
(\hat{t_n})^{i+1} =& {z_n}^{i+1}+\frac{(\theta^t_n)^i-\mu_t}{s}, \label{equ5-B-8}
\end{align}
where $\mu_f$ can be obtained by the bisection search method over $(0, \mu_f^{\star})$, until $\sum_{n\in \mathcal{N}_1}  x_n e^{\hat{f_n^e}} \le f^{max}$ satisfies. $\mu_f^{\star}$ is a sufficiently large value. It is because when $\mu_f \ge 0$, $(\hat{f_n^e})^{i+1}$ is non-increasing. Similarly, $\mu_t$ can be obtained by the bisection search method over $(0,\mu_t^{\star})$, where $\mu_t^{\star}$ is a sufficiently large value, until $\sum_{n\in \mathcal{N}_1} x_n e^{\hat{t_n}} \le 1$ satisfies.

\subsubsection{Step 3} Multipliers update. In this step, we update the multipliers $\bm{\theta}$ using the obtained global variables $\bm{v}$ and local variables $\bm{u}$. The updated method is,
\begin{align}
(\theta_n^f)^{i+1} =& (\theta_n^f)^{i}+s(y_n^{i+1}-(\hat{f_n^e})^{i+1}), \label{equ5-B-9}\\
(\theta_n^t)^{i+1} =& {z_n}^{i+1}+s(z_n^{i+1}-(\hat{t_n})^{i+1}), \label{equ5-B-10}
\end{align}

Repeat the above three steps until the cost function no longer decreases. The cost function is $\mathcal{F}^i=\sum_{n \in \mathcal{N}}[(1-x_n^i)\hat{\mathcal{F}_{0,n}}((\hat{M_n})^i,(\hat{f_n^{md})^i}) \nonumber +x_n^i \hat{\mathcal{F}_{1,n}}((\hat{M_n})^i,(\hat{f_n^e})^i,(\hat{t_n})^i)]$. We summarize solving steps of the ADMM algorithm as Algorithm \ref{algorithm4}.

\textcolor{black}{As a distributed iterative algorithm, the ADMM-based scheme performs iterations between devices and BS rather than locally, enabling online optimization during the recognition process. In each iteration, $\bm{u}^i$ is calculated locally and sent to the MEC. After receiving $\bm{u}^i$ from all devices, the MEC updates $\bm{v}^i$ and $\bm{\theta}^i$, and sends them to the device to complete an iteration. Therefore, the iteration of the ADMM algorithm is an online convergence process that can adapt to slight changes in the channel.}

\begin{algorithm}[t]
	\caption{Algorithm 4: ADMM-Based Algorithm}
	\label{algorithm4}
	\KwIn{Parameters corresponding to the problem $\mathcal{P}1$.}
	\KwOut{$\{x_n,{M_n},{f_n^{md}},{f_n^{e}},{t_n}\}$}
	\BlankLine

    Initialize $i=0$, $\{\bm{u}^i,\bm{v}^i,\bm{\theta}^i\}=0$, $s=0.5$, $\mu_f^{\star}=\mu_t^{\star}=10^{6}$, $\delta=10^{-4}$;

    \Repeat{$|\mathcal{F}^i-\mathcal{F}^{i+1}|<\delta$} {
    \ForEach{$n \in \mathcal{N}$}{ 
    Update $\bm{u}^{i+1}$ by solving \eqref{equ5-B-5} and choose smaller results;
    }
    \ForEach{$n \in \mathcal{N}$}{
    Update global variables $\bm{v}^{i+1}$ using \eqref{equ5-B-7} and \eqref{equ5-B-8};
    }
    \ForEach{$n \in \mathcal{N}$}{
    Update multipliers $\bm{\theta}^{i+1}$ using \eqref{equ5-B-9} and \eqref{equ5-B-10};
    }
    $i=i+1$;
    }
    
    $M_n=e^{\hat{M_n}}$, ${f_n^{md}}=e^{\hat{f_n^{md}}}$, ${f_n^{e}}=e^{\hat{f_n^{e}}}$, $t_n=e^{\hat{t_n}}$;
    
    \Return {$\{x_n,{M_n},{f_n^{md}},{f_n^{e}},{t_n}\}$}.
\end{algorithm}

\subsection{Algorithm Computational Complexity Analysis}

In this part, we analyze the computational complexity of proposed algorithms. First, the complexity of solving problem $\mathcal{P}_{\mathcal{N}_0}$ is $O(|\mathcal{N}_0|)$. Second, as mentioned above, the complexity of Algorithm \ref{algorithm1} is $O(\prod_{n \in \mathcal{N}_1} |\mathcal{M}_n^{opt}|)$, and the complexity of Algorithm \ref{algorithm2} is $O((3|\mathcal{N}_1|)^{3.5})$ by the interior-point method according to \cite{interior-point}. When we use Algorithm \ref{algorithm1} for solving $\mathcal{P}_{\mathcal{N}_1}$ and use Search-based algorithm for optimizing offloading policy, the computational complexity is $O(2^N \prod_{n \in \mathcal{N}} |\mathcal{M}_n^{opt}|)$. When we use Algorithm \ref{algorithm1} for solving $\mathcal{P}_{\mathcal{N}_1}$ and use Algorithm \ref{algorithm3} for optimizing offloading policy, the computational complexity is $O(N \prod_{n \in \mathcal{N}} |\mathcal{M}_n^{opt}|)$. In addition, the computational complexity of Algorithm \ref{algorithm2} for solving $\mathcal{P}_{\mathcal{N}_1}$ and Algorithm \ref{algorithm3} for optimizing offloading policy is $O(N^{4.5})$. For the ADMM-based algorithm, as the complexity of each steps is $O( \mathcal{N})$, the overall complexity of one iteration is $O( \mathcal{N})$.

\section{NUMERICAL RESULTS} \label{Sec6}
In this section, we evaluate the performance of the proposed algorithms via simulations. For all the simulation results, unless specified otherwise, we set the downlink bandwidth as $B_w=5$ MHz and the power spectral as $N_0=-174$ dBm/Hz \cite{EI-accuracy1}. According to \cite{EI-energy1}, the path loss is modelled as $PL=128.1+37.6\log_{10}(D)$ dB, where $D$ is the distance between the device and the BS in kilometres. Devices randomly distributed in the area within  [500m 500m]. The computational resource of the MEC server and devices are set to be 1.8 GHz and 22 GHz, respectively. The recognition accuracy requirement and the maximum number of input video frames are set to $\alpha_n=0.8
6$ and $M^{max}_n=16$, respectively. The coefficient $\kappa$ is determined by the corresponding device and is set to be $10^{-28}$ in this paper according to \cite{EI-accuracy1}. The size of the input video is $112 * 112 * M_n$. In addition, the coefficient of computational complexity $\rho$ is set to be 0.12 cycle/MAC, which is obtained through several experiments in Sec.\ref{S5-1}. Weights $\beta_1, \ \beta_2, \ \beta_3$ are set to be 0.2, 0.2, 0.6, respectively.



\subsection{Model Verification}\label{S5-1}

\begin{figure}[tb]
\centering 
\includegraphics[height=2.0in,width=2.6in]{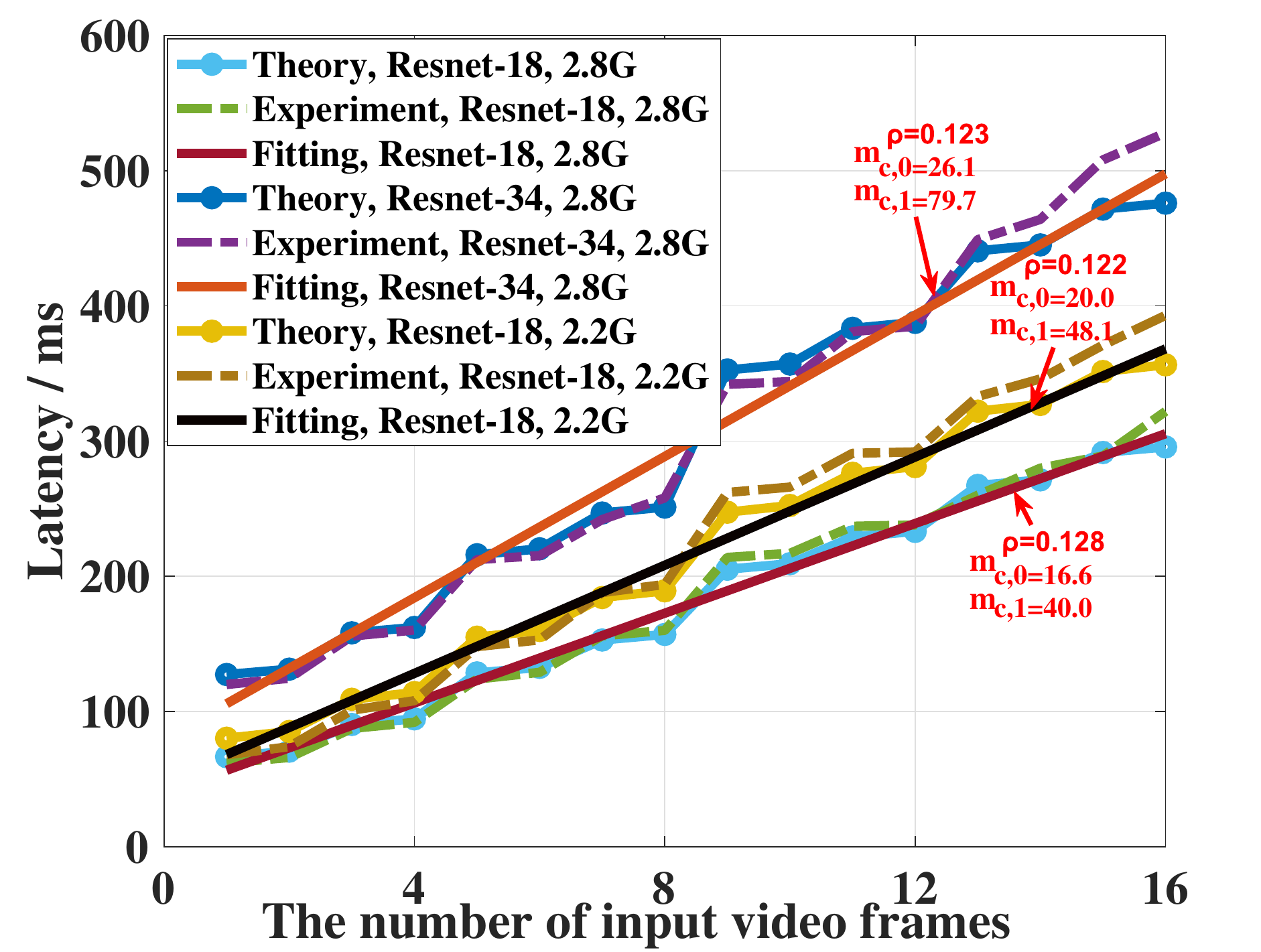} \vspace{-2 mm}
\caption{The theoretical delay curve, the experimental delay curve and the fitted curve corresponding to the experimentalal delay. Resnet-18 and Resnet-34 are two classic neural network architectures. The frequency of the CPU is 2.8G and 2.2G.}
\label{fitting1}
\vspace{-2 mm}
\end{figure}

\begin{figure}[tb]
\centering 
\includegraphics[height=2.0in,width=2.6in]{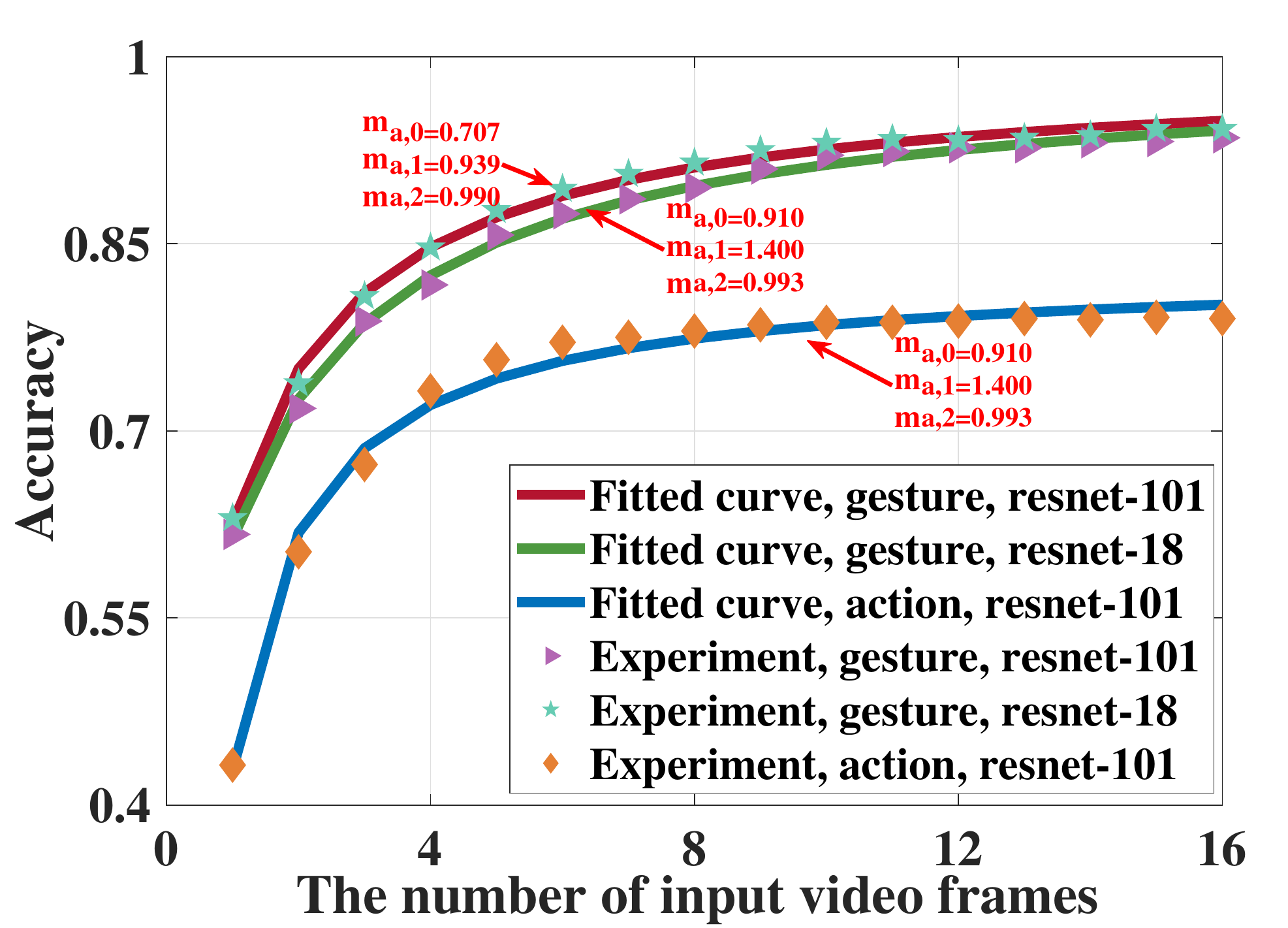}
\caption{The experimental and fitted curves of gesture recognition task and action recognition task.}
\label{fitting2} 
\vspace{-2 mm}
\end{figure}

First, we obtain the complexity coefficient through experimental measurement. The calculation method of the computational complexity coefficient is as follows. First, calculate the MACs of the DNN model when the number of input video frames is different, recorded as $\{C\}$. We use the Flops Counter tool \cite{flop} for MACs calculation. Second, execute 100 times of inference tasks with a different number of input video frames, and record the average inference delay as $\{t\}$. Finally, calculate the coefficients between the inference delay and MACs by $\rho=\frac{sum(\{C\})}{sum(\{t\})}$. We use Intel(R) Xeon(R) E5-2630 CPU for testing. We use the Resnet-18 and the Resnet-34 for testing and limit the maximum frequency of the CPU to 2.8G and 2.2G. Fig. \ref{fitting1} shows the theoretical (MAC-based) and experimental delay curves and the fitted curve corresponding to the experimental delay. We can observe from Fig. \ref{fitting1} that the theoretical delay is similar to the experimental delay, proving that MACs can be modelled as computational complexity. We also find that the linear fitted curve can approximately represent the computational complexity with 9 ms root mean square error (RMSE) for Resnet-18 and 2.8G, 17 ms RMSE for Resnet-34 and 2.8G, and 11 ms RMSE for Resnet-18 and 2.2G. The inference delay is associated with the number of input frames, DNN model’s architecture and the device’s capabilities. In addition, the computational complexity coefficients under the three conditions are 0.128, 0.122, and 0.123, respectively. Therefore, in following experiments, we set $\rho=0.12$ cycle/MAC.

We select the gesture and action recognition tasks to verify the accuracy model. We use the Jester datasets \cite{Jester}, the largest publicly available hand gesture dataset, to test the gesture recognition task. For the action recognition task, we use Kinetics-400 datasets \cite{kinetics}. We choose Resnet-18 and Resnet-101 for testing. As shown in Fig. \ref{fitting2}, Under different tasks and different network models, the accuracy curve all conforms to the characteristics of a non-decreasing function. \textcolor{black}{What's more, as the number of input frames increases, the performance gain of accuracy will gradually decrease. This is because the information gain introduced in the temporal domain decreases when the number of input frames increases.} The fitted curve can approximately represent the relationship between the accuracy and the number of input frames. In the gesture recognition task with the Resnet-101 model, the gesture recognition task with the Resnet-18 model, and the action recognition task with the Resnet-101 model, the RMSE are 0.0054, 0.0048 and 0.0095, respectively. We take the Resnet-18 and the gesture recognition task as examples for the following experiments.

\subsection{Simulation Results of Average Cost}

In this section, we compare proposed schemes and some baseline schemes. We run 100 tests and can calculate the average cost of each device and the average running time of each test. \textcolor{black}{We compare the following schemes.}
\color{black}
\subsubsection{\textbf{Search+Search}}We use the Search-based algorithm to solve $\mathcal{P}_{\mathcal{N}_1}$ and use the heuristic algorithm to optimize offloading policy.

\subsubsection{\textbf{Search+Heuristic}}We use the Search-based algorithm to solve $\mathcal{P}_{\mathcal{N}_1}$ and use the Search-based algorithm to optimize offloading policy.

\subsubsection{\textbf{GP+Heuristic}} We use the GP-based algorithm to solve $\mathcal{P}_{\mathcal{N}_1}$ and use the Channel-Aware heuristic algorithm to optimize offloading policy. 

\subsubsection{\textbf{ADMM}} We use the ADMM-based algorithm to solve the original problem.

\subsubsection{\textbf{CCCP} \cite{CCCP}} We use the concave-convex procedure (CCCP) algorithm to decide whether to offload inference tasks to edge servers. Then we use \textit{Theorem 1} and the GP-based algorithm for resource allocation.

\subsubsection{\textbf{Random}} All inference tasks are randomly executed on local or the edge server. We use \textit{Theorem 1} and the GP-based algorithm for resource allocation.

\subsubsection{\textbf{Local}} All inference tasks are executed locally. We use \textit{Theorem 1} for local resource allocation.

\subsubsection{\textbf{Edge}} All inference tasks are executed on the edge server. We use the GP-based algorithm for resource allocation

\color{black}

\begin{figure}[tb]
\centering 
\includegraphics[height=2.0in,width=2.6in]{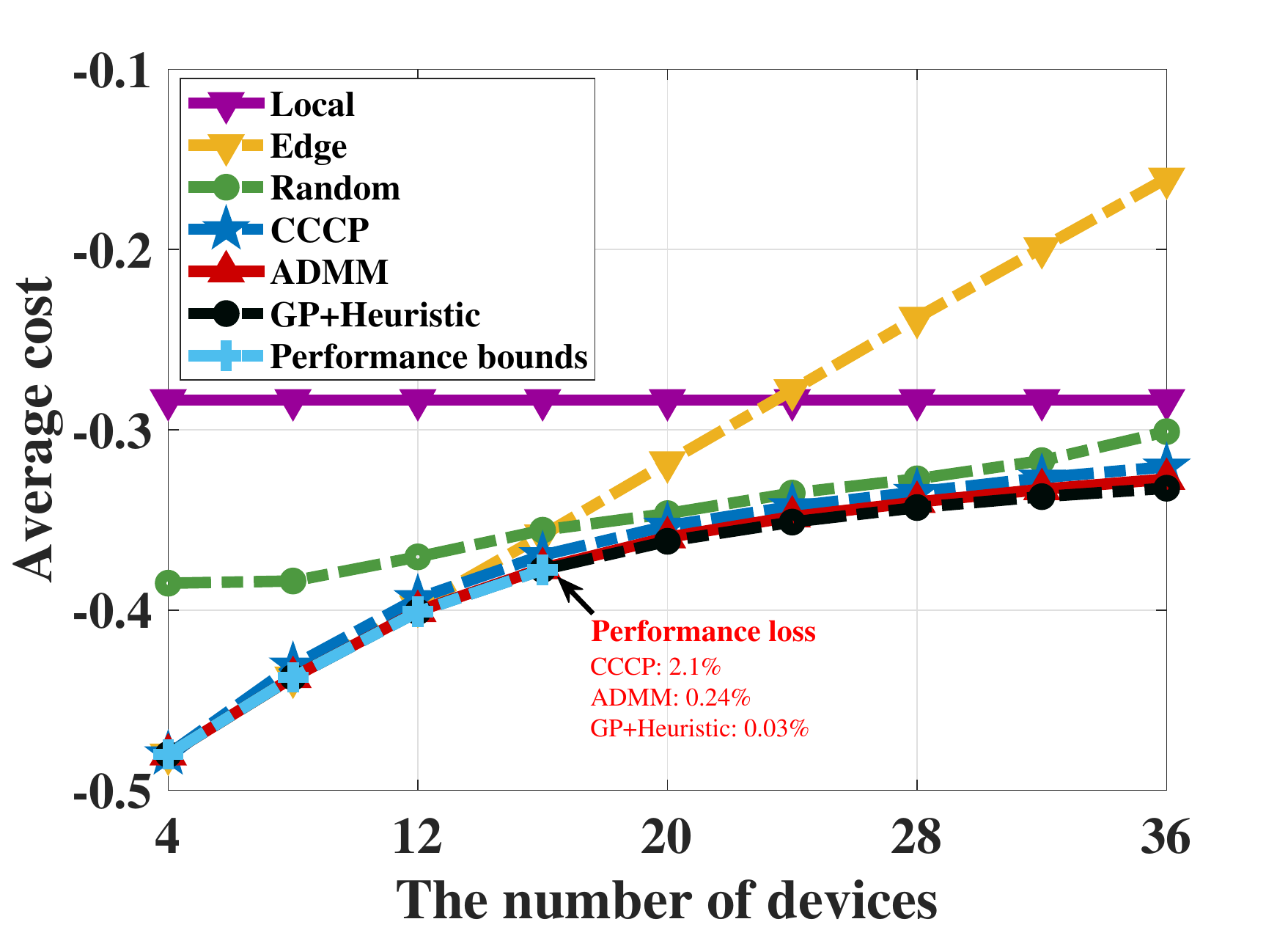}
\vspace{-1 mm}
\caption{\textcolor{black}{The average cost of proposed schemes and baseline schemes under a different number of devices.}}
\label{result_cost} 
\vspace{-2 mm}
\end{figure}

\color{black}

\textcolor{black}{In Fig. \ref{result_cost}, we plot the average cost of different schemes under different devices. The Search+Heuristic scheme and Search+Search scheme have the same performance, representing the performance bounds. When the number of devices exceeds 16, the performance bounds are not shown due to their unacceptable computational complexity. It can be seen from Fig. \ref{result_cost} that the proposed schemes are better than the baseline schemes. Compared with the performance bounds, the performance of the GP+Heuristic scheme has a slight decrease due to the relaxation of the accuracy function ${\Phi}(M_{n})$. The performance of the ADMM scheme is worse than that of the GP+Heuristic scheme, and is better than that of the CCCP scheme. For example, when the number of devices is 16, the CCCP, ADMM, and GP+Heuristic schemes have performance losses of 2.1\%, 0.24\%, and 0.03\%, respectively, compared with performance bounds.} Moreover, when the number of devices is less than 8, the cost of the scheme that executes tasks only at the edge is almost equal to the cost of the proposed GP+Heuristic scheme. It is because all devices can benefit from performing inference on the edge server when the number of devices is small. If the inference task is only executed locally, the average cost of the device will not change because the local resources among the equipment do not affect each other.

\begin{figure}[tb]
\centering 
\includegraphics[height=2.0in,width=2.6in]{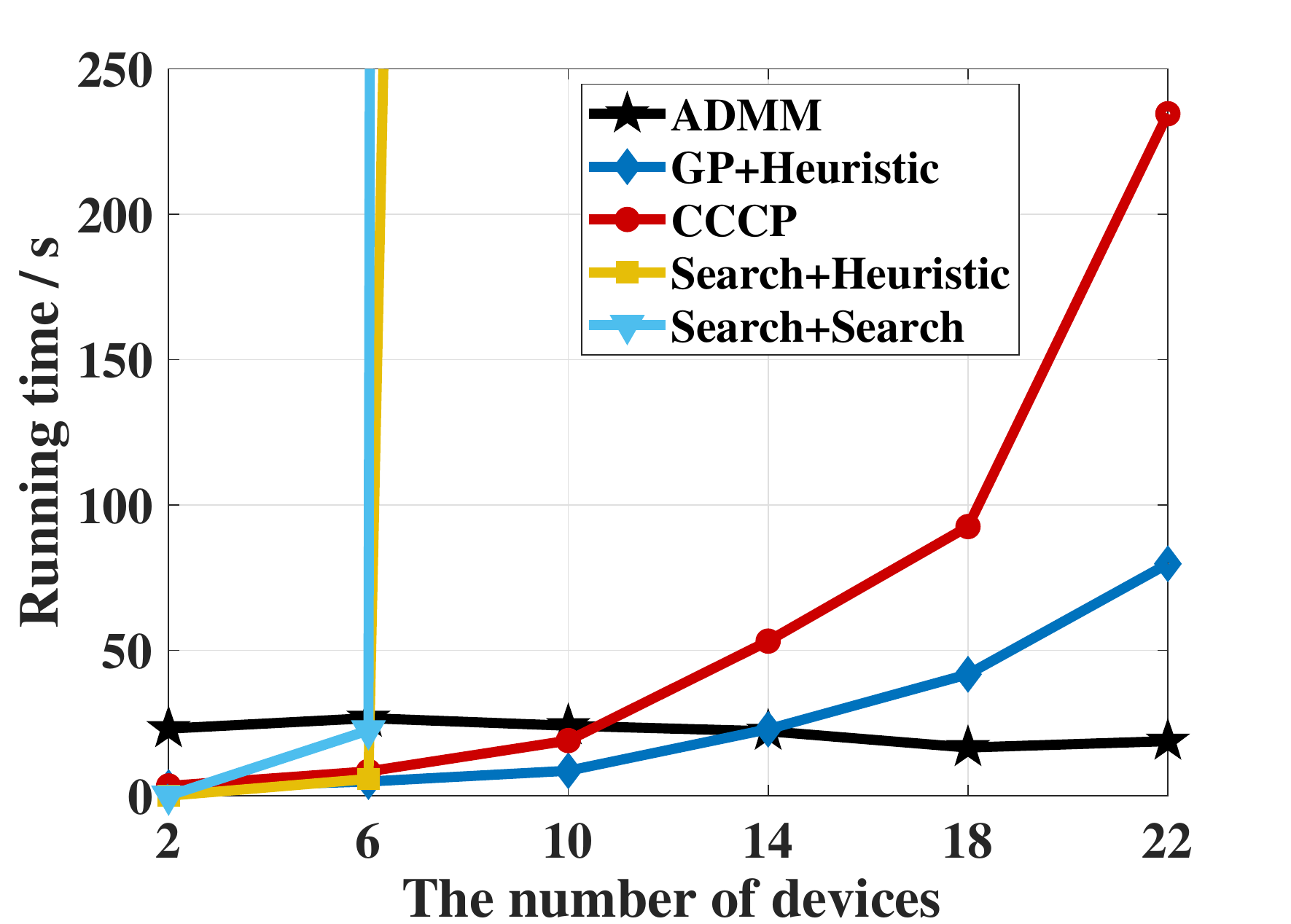} 
\vspace{-1 mm}
\caption{The average running time of proposed algorithms under a different number of devices.}
\label{result_runningtime} 
\vspace{-5 mm}
\end{figure}

\textcolor{black}{In Fig. \ref{result_runningtime}, we plot the average running time of different schemes under different devices.} When the number of devices exceeds 6, the running time of the Search+Heuristic and Search+Search scenarios becomes unacceptable. The GP+Heuristic scheme improves the solution efficiency. \textcolor{black}{The running time of GP+Heuristic is shorter than that of CCCP scheme.} However, the complexity of the solution remains unsatisfactory as the number of devices increases. \textcolor{black}{As for the ADMM-based scheme, since the ADMM-based algorithm is a distributed algorithm and the complexity of updating global variables is much smaller than that of updating local variables, we only consider the average running time for each device. The average running time of the ADMM-based scheme does not improve as the number of devices increases. It is worth noting that in the ADMM-based scheme, the iteration stops when $|\mathcal{F}^i-\mathcal{F}^{i+1}|<\delta$, where $\delta=10^{-5}$. Threshold-based stopping conditions result in a different number of iterations in different cases. When the number of devices is different, the average number of iterations is also different, resulting in different running times. Therefore, the average running time of 18 devices is shorter than that of 14 and 22 devices.}

\begin{figure}[tb]
\centering 
\includegraphics[height=2.0in,width=2.6in]{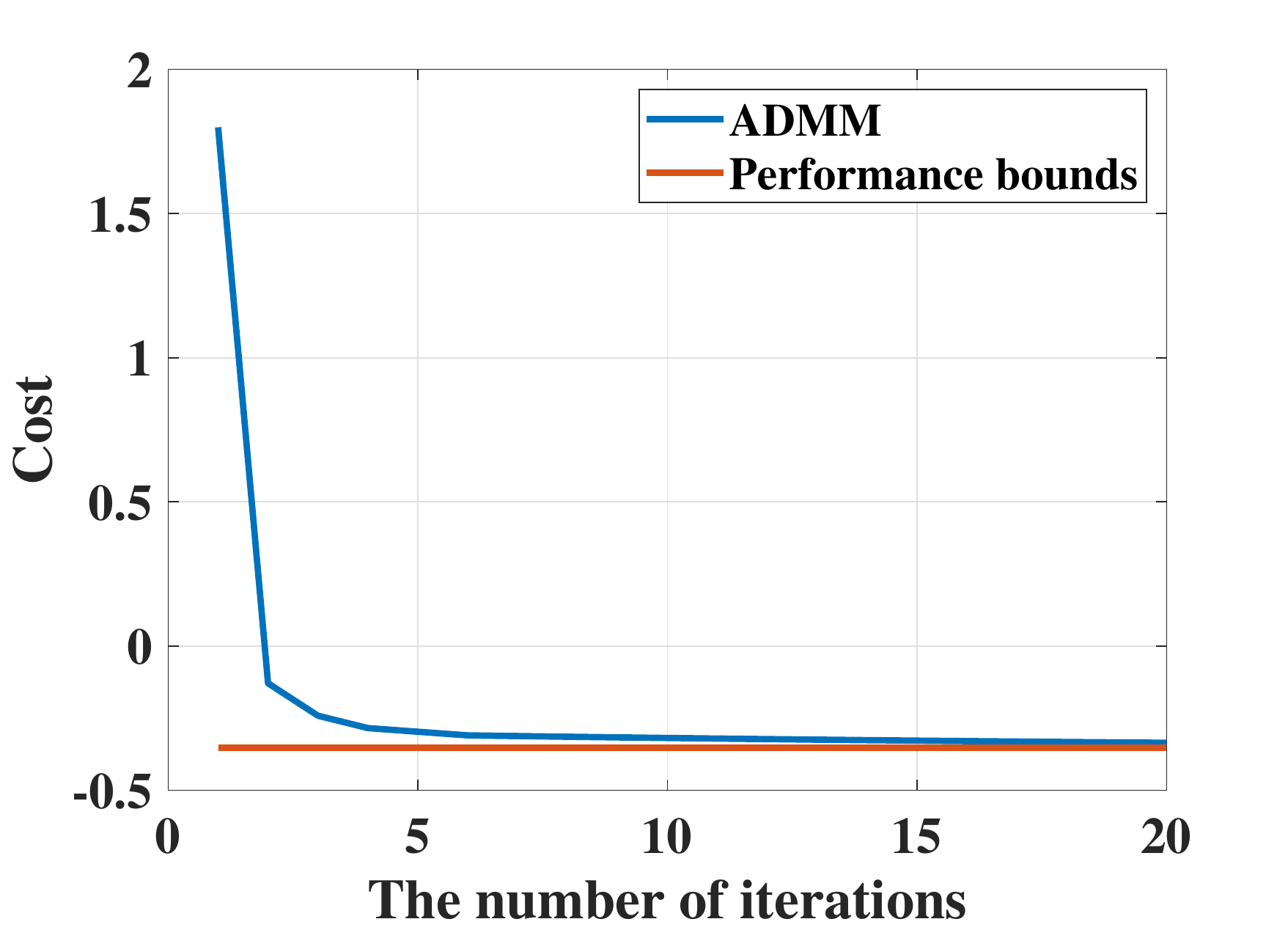} 
\vspace{-1 mm}
\caption{The curve corresponding to the cost function and the number of iterations.}
\label{ADMM} 
\vspace{-2 mm}
\end{figure}

\textcolor{black}{ Assuming that the ADMM-based scheme iterates once every time an inference task is performed, we plot the curve corresponding to the cost function and the number of iterations. As shown in Fig.7, the ADMM-based scheme can converge to acceptable performance after completing 3-5 iterations. As the number of iterations increases, the performance will be closer to the optimal performance. It shows that the ADMM algorithm can converge through online iterations.  We also test the running time per iteration on each device, and it takes an average of about 278ms.}

\subsection{Simulation Results of Delay, Energy, and Accuracy}

\begin{figure}[t]
\subfigure[Different number of devices]{
\begin{minipage}[t]{1\linewidth}
\centering 
\includegraphics[height=1.8in,width=2.4in]{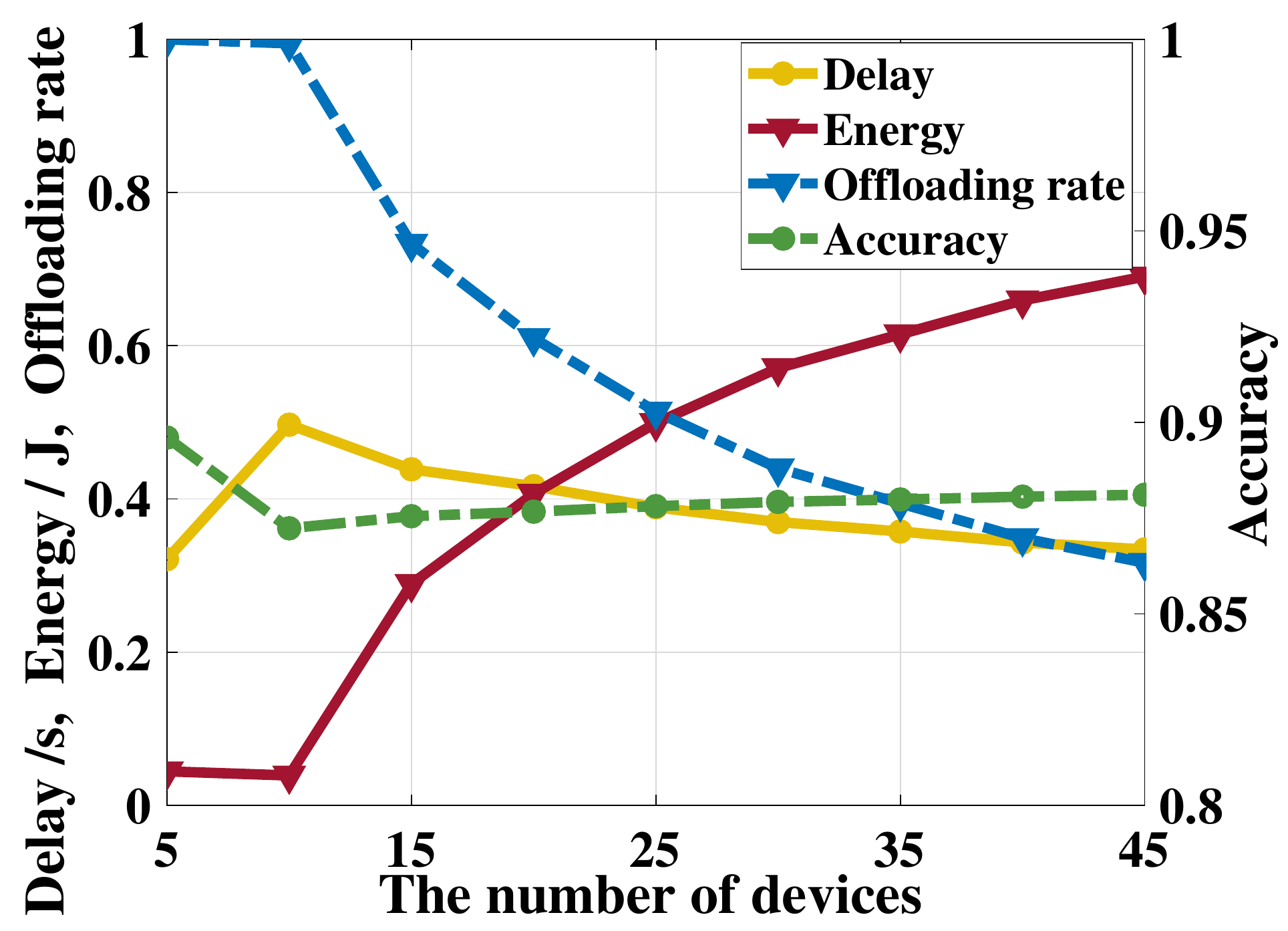} 
\label{result-device} 
\end{minipage}}
\subfigure[Different bandwidth]{
\begin{minipage}[t]{1\linewidth}
\centering 
\includegraphics[height=1.8in,width=2.4in]{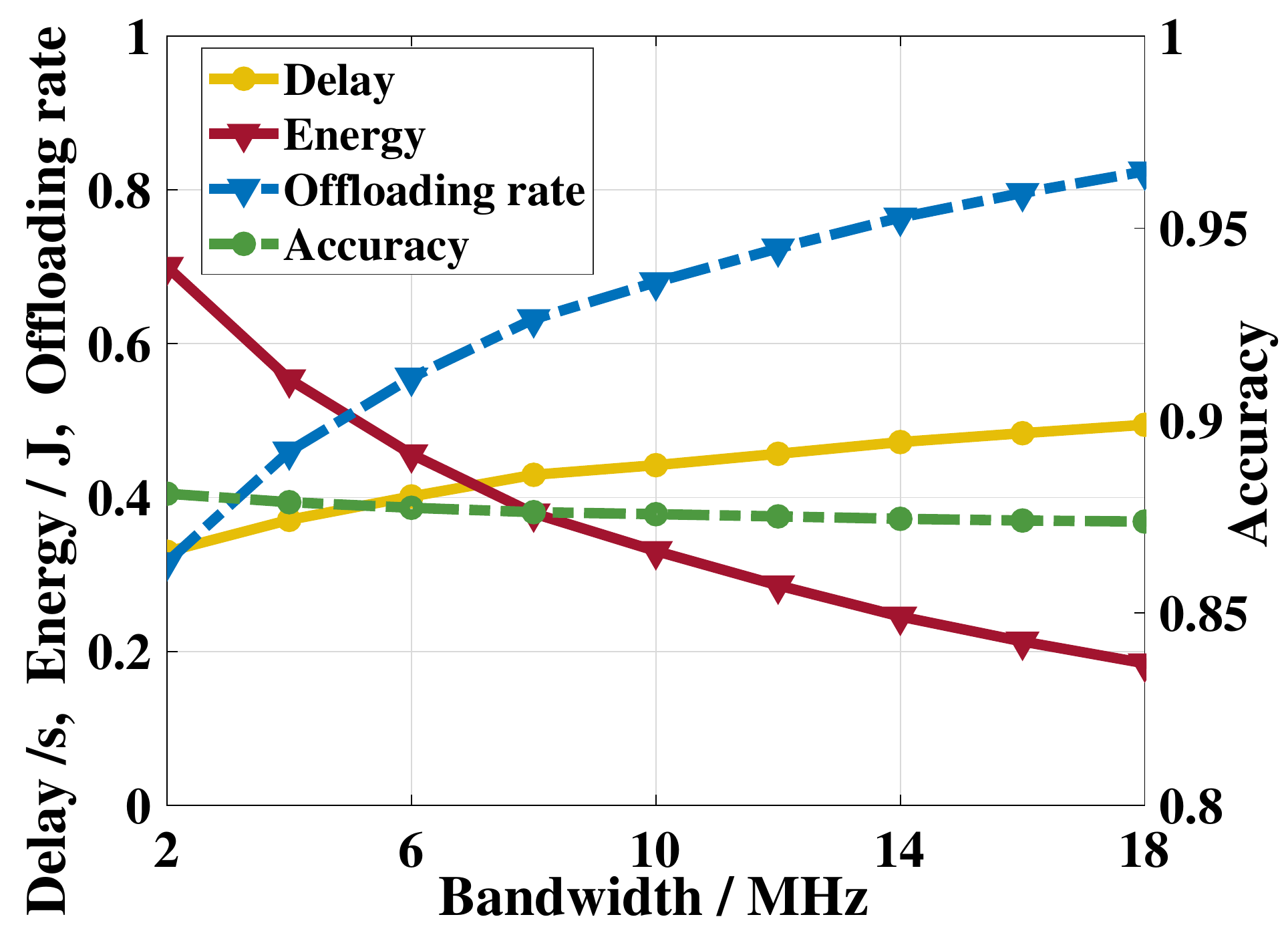} 
\label{result-BW} 
\end{minipage}}
\subfigure[Different edge computing resource]{
\begin{minipage}[t]{1\linewidth}
\centering 
\includegraphics[height=1.8in,width=2.4in]{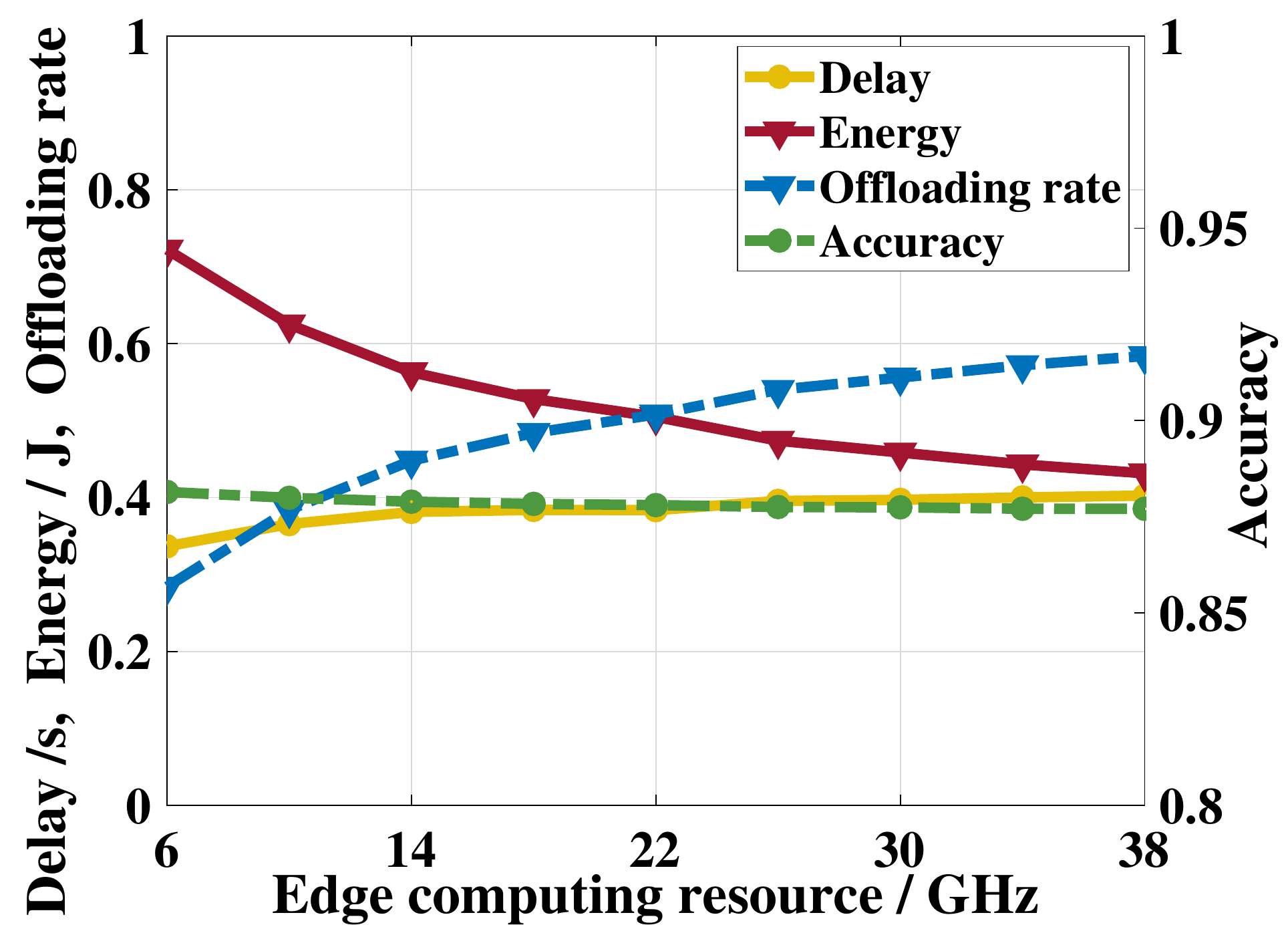} 
\label{result-EC} 
\end{minipage}}
\caption{The average delay, energy, offloading rate, and accuracy under different numbers of devices, different bandwidths, and different edge computing resources.}
\label{fig:result-C1}
\vspace{-4 mm}
\end{figure}

\begin{table}[t]\footnotesize 
	\centering 
	\caption{Delay, energy consumption, and accuracy of local devices and edge devices} 
	\label{table2}
  \renewcommand\arraystretch{1.5}
	\begin{tabular}{|c|c|c|} 
    \bottomrule
		& Local devices  & Edge devices \\ 
		\hline \bottomrule
		Number of devices & 12.3& 12.7\\
		\hline
		Average delay & 0.24 s& 0.52 s\\
		\hline
		Average energy &1.00 J& 0.025 J\\
		\hline
		Average accuracy &0.886 & 0.866\\
		\hline \bottomrule
	\end{tabular}
	\vspace{-4 mm}
\end{table}

This section compares the average delay, energy consumption, accuracy, and the offloading rate (the proportion of devices that perform inference on the edge server). We consider the different number of devices, bandwidths, edge computing resources, and weights $\beta_1$, $\beta_2$, $\beta_3$. We use the GP+Heuristic scheme for testing. Table. \ref{table2} shows a comparison of devices that finish inference locally and devices that finish inference at the edge under default experimental settings. On average, 12.7 devices choose to offload to the edge server to perform inference. Compared with edge devices, local devices have a lower delay and higher accuracy but have greater inference energy consumption.

Fig. \ref{fig:result-C1} shows the average delay, energy, accuracy, and offloading rate under different numbers of devices, different bandwidths, and different edge computing resources. In Fig. \ref{result-device}, we plot results with different numbers of devices. \textcolor{black}{As shown in Fig. \ref{result-device}, when the number of devices is small (less than 10), all devices offload the task to the edge server (the offloading rate is equal to 1). For edge devices, all delay comes from transmission delay and the edge inference delay, and all energy consumption comes from transmission energy. With the number of devices increasing, communication resources and the edge server's computation resources are shared by more devices, decreasing the number of input frames $M_n$. A decrease in the number of input frames results in a decrease in accuracy. Then as $M_n$ decreases, the transmission data size decreases, and the transmission energy decreases. Meanwhile, Competition from more devices leads to increased delays. Therefore, when the number of devices is small (less than 10), with the number of devices increasing, the average delay increases, the average accuracy and the average energy consumption decrease. When the number of devices exceeds 10, the average energy consumption and accuracy increase, and the average delay and offload rate gradually decrease. Considering different bandwidths and different edge computing resources, we plot Fig. \ref{result-BW} and Fig. \ref{result-EC}. In Fig. \ref{result-BW} and Fig. \ref{result-EC}, as the bandwidth and edge computing resource increase, devices will be more inclined to offload computing to the edge, which increases the offloading rate. According to Table. \ref{table2}, when  $\beta_1, \ \beta_2$ and $ \beta_3$ are fixed, edge devices have lower energy consumption, lower accuracy and higher delay. More edge devices mean a greater delay and lower power consumption. Meanwhile, when the bandwidth increases, since the edge computing resources are fixed, the number of video frames will decrease to reduce edge computing overhead, resulting in a decrease in accuracy. The same conclusion can also be obtained when edge computing resources increase. Therefore, with the increase of bandwidth and edge computing resources, more edge devices lead to increased delay and decreased energy and accuracy.}

\begin{figure}[tb]
\centering 
\includegraphics[height=2.2in,width=3.0in]{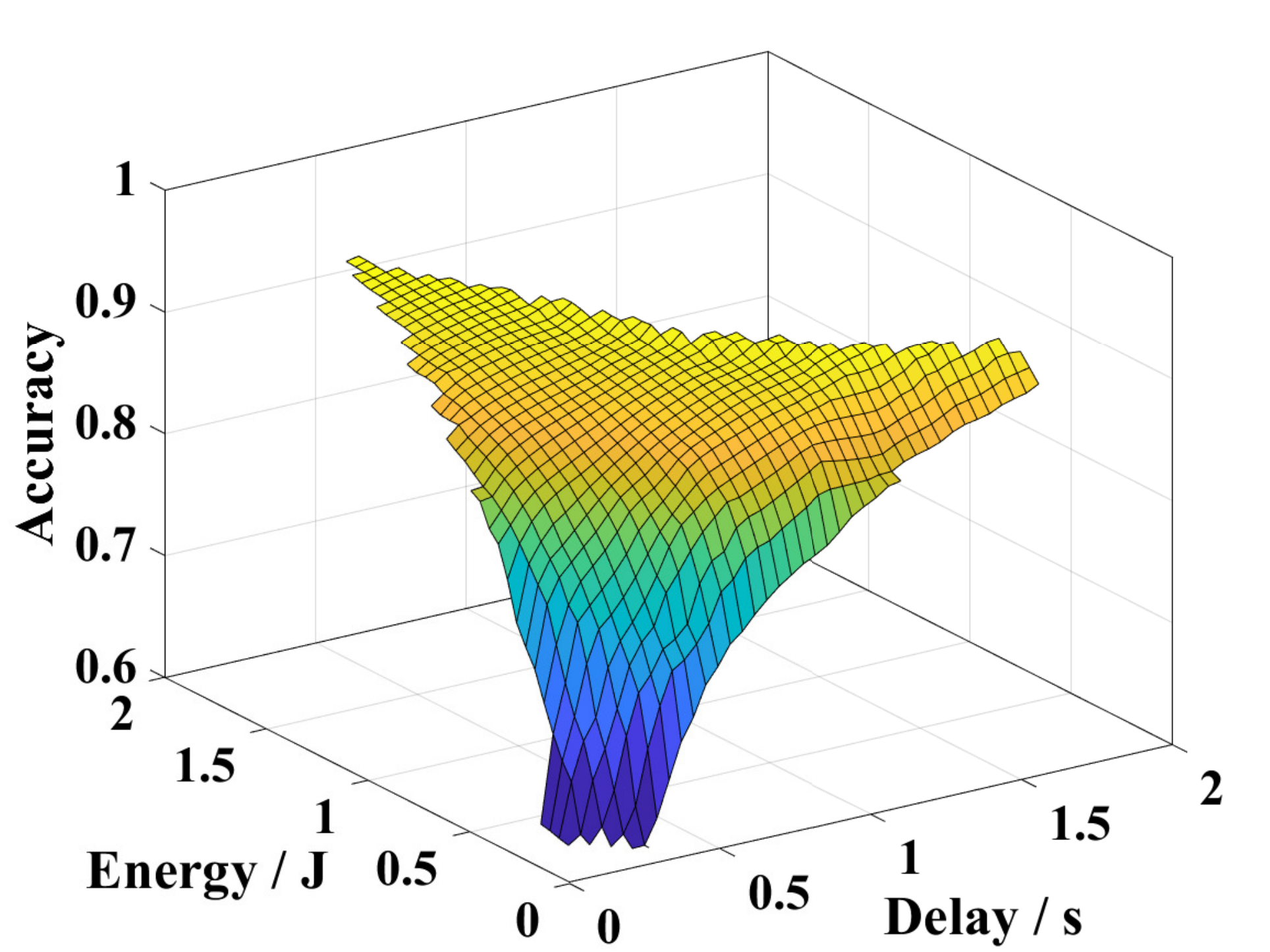}
\caption{The relationship between the delay, energy consumption, and accuracy.}
\label{fig-result-C2}
\vspace{-4 mm}
\end{figure}

We set the minimum number of input frames $M_n^{min}=1$. We use different weights, $\beta_1, \ \beta_2, \ \beta_3$ to study the trade-off relationship between the average delay, energy consumption, and accuracy. The constraint is $\beta_1+ \beta_2+ \beta_3=1$. The performance of the trade-off surface is obtained by the GP+Heuristic scheme. Fig. \ref{fig-result-C2} shows the delay, energy consumption, and accuracy are mutually limited. Higher energy consumption leads to higher accuracy when the delay is constant. From another perspective, in order to improve the accuracy, it is necessary to sacrifice the performance of delay and energy consumption. In addition, with the same accuracy, according to Table. \ref{table2}, higher energy consumption will make the device more inclined to execute inference tasks locally, and the delay decreases.

\section{Conclusion} \label{Sec7}
This paper considers optimizing video-based AI inference tasks in a multi-user MEC system. An MINLP is formulated to minimize the total delay and energy consumption, and improve the total accuracy, with the constraint of computation and communication resources. A MAC-based computational complexity model is introduced to model the calculation delay, and a simple approximate expression is proposed to simplify the problem. We also propose a general accuracy model to characterize the relation between the recognition accuracy and the number of input frames. After that, we first assume that the offloading decision is given and decouple the original problem into two sub-problems. The first sub-problem is to optimize the resources of the devices that complete the DNN inference tasks locally. We derive the closed-form solution to this problem. The second sub-problem is optimizing the devices' resources that offload the DNN inference tasks to the edge server. We propose the Search-based and GP-based algorithm to solve the second sub-problem. For the problem of offloading decision optimization, we propose the Channel-Aware heuristic algorithm. We also propose a distributed algorithm based on ADMM. The ADMM-based algorithm reduce computational complexity at the cost of an acceptable performance loss. Numerical simulation and experimental results demonstrate the effectiveness of the proposed algorithm. We also provide a detailed analysis of the delay, energy consumption, and accuracy for different device numbers, bandwidths and edge computing resources.


%

\appendices
\section{Proof of Theorem 1}
The partial derivative of $\mathcal{F}_{\mathcal{P}_{\mathcal{N}_0}}$ with respect to $f_n^{md}$ is,
\begin{align}
\frac{\partial \mathcal{F}_{\mathcal{P}_{\mathcal{N}_0}} }{\partial f_n^{md}} = - \beta_1\frac{\rho C(M_{n})}{f_n^{md2}}+2\beta_2 \kappa\rho C(M_n)f_n^{md}, \label{equ-appendices-A-1}
\end{align}
By setting $\frac{\partial \mathcal{F}_{\mathcal{P}_{\mathcal{N}_0}} }{\partial f_n^{md}}=0$, we have,
\begin{align}
f_n^{md} = \sqrt[3]{(\frac{\beta_1}{2\beta_2\kappa})}, \label{equ-appendices-A-2}
\end{align}
Therefore, $f_n^{md}$ decreases monotonically in the interval $(-\infty, \sqrt[3]{(\frac{\beta_1}{2\beta_2\kappa})})$ and increases monotonically in the interval $( \sqrt[3]{(\frac{\beta_1}{2\beta_2\kappa})},+\infty)$. Considering the value range of $f_n^{md}$, the optimal solution can be given by,
\begin{align}
f_n^{md*} = & \ \textrm{min}\{\sqrt[3]{(\frac{\beta_1}{2\beta_2\kappa})}, f_n^{max}\} \label{equ-appendices-A-3}
\end{align}

Then we analyze $M_n$. The partial derivative of $\mathcal{F}_{\mathcal{P}_{\mathcal{N}_0}}$ with respect to $M_n$ is,
\begin{align}
\frac{\partial \mathcal{F}_{\mathcal{P}_{\mathcal{N}_0}} }{\partial M_n} =  \frac{ \beta_1 \rho m_{c,0}}{f_n^{md}}+\beta_2 \kappa\rho m_{c,0}f_n^{md2}-\frac{\beta_3 m_{a,0}}{(M_n+m_{a,1})^2}, \label{equ-appendices-A-4}
\end{align}
By setting $\frac{\partial \mathcal{F}_{\mathcal{P}_{\mathcal{N}_0}} }{\partial M_{n}}=0$, we have,
\begin{align}
M_n = \sqrt{\frac{\beta_3 m_{a,0}}{\frac{ \beta_1 \rho m_{c,0}}{f_n^{md}}+\beta_2 \kappa\rho m_{c,0}f_n^{md2}}}-m_{a,1}, \label{equ-appendices-A-5}
\end{align}
Considering the value range of $M_n$, the optimal solution can be given by,
\begin{align}
M_n^{*} = & \ \textrm{min} \{\textrm{max}\{ \sqrt{\frac{\beta_3 m_{a,0}}{\frac{ \beta_1 \rho m_{c,0}}{f_n^{md}}+\beta_2 \kappa\rho m_{c,0}f_n^{md2}}} \nonumber \\
&-m_{a,1}, M^{min}_n\},M^{max}_n\} \label{equ-appendices-A-6}
\end{align}
which completes the proof.

\section{Proof of Theorem 2}
According to the KKT conditions, we can obtain the following necessary and sufficient conditions,
\begin{flalign}
& \frac{\partial \mathcal{L}_{\mathcal{P}_{\mathcal{N}_1}} }{\partial f_n^{e*}} = -\frac{\beta_1\rho C(M_{n}^{*})}{f_n^{e*2}}+u_1^{*}=0 ,\ f_n^{e*}>0, \label{equC-1}  \\
& \frac{\partial \mathcal{L}_{\mathcal{P}_{\mathcal{N}_1}} }{\partial t_n^{*}} = -\frac{\beta_1 M_n^{*} d}{R_n t_n^{*2}}+u_0^{*}=0,  t_n^{*}>0,\ \label{equC-2}  \\
& \mu_0^{*} (\sum_{n\in \mathcal{N}^{*}} t_n^{*} - 1)=0,  \label{equC-3} \\
& \mu_1^{*} (\sum_{n\in \mathcal{N}^{*}} f_n^{e*} - f^{max})=0, \label{equC-4} \\
& \mu_0^{*}, \mu_1^{*} \ge 0. \label{equC-5} 
\end{flalign}

Because $\frac{\beta_1\rho C(M_{n}^{*})}{f_n^{e*2}}$ and $\frac{\beta_1M_n^{*} d}{R_n t_n^{*2}}$ are positive, $\mu_0^{*}$ and $\mu_1^{*}$ are also positive. We can obtain,
\begin{flalign}
& \sum_{n\in \mathcal{N}} f_n^{e*} - f^{max}=0,  \label{equC-6} \\
& \sum_{n\in \mathcal{N}} t_n^{*} - 1=0, \label{equC-7} \\
& f_n^{e*}=\sqrt{\frac{\beta_1\rho C(M_{n}^{*})}{R_n \mu_1^{*}}}, \label{equC-8} \\
& t_n^{*}=\sqrt{\frac{\beta_1 M_n^{*} d}{R_n\mu_0^{*}}}. \label{equC-9} 
\end{flalign}

Combining \eqref{equC-6} and \eqref{equC-8}, we can get the expression of $f_n^{e*}$ corresponding to $M_n^{*}$,
\begin{align}
f_n^{e*} = & \ \frac{f^{max} \sqrt{C(M_n^{*})}}{\sum\limits_{i\in \mathcal{N}_1} \sqrt{C(M_i^{*})}}.
\end{align}
Similarly, combining \eqref{equC-7} and \eqref{equC-9}, we can get the expression of $t_n^{*}$ corresponding to $M_n^{*}$,
\begin{align}
t_n^{*} =  \frac{\sqrt{\frac{M_n^{*}}{R_n}}}{\sum\limits_{i\in \mathcal{N}_1} \sqrt{\frac{M_i^{*}}{R_i}}},
\end{align}
which completes the proof.


\bibliographystyle{IEEEtran}

\bibliography{reference}
\end{document}